\documentclass[twocolumn]{aastex62}
\usepackage{graphicx}
\usepackage{subfig}

\newcommand{\OIII}{{\ion{O}{3}}}

\newcommand{\kms}{km s$^{-1}$}

\newcommand{\ergs}{erg s$^{-1}$} 

\usepackage{subfig}
\usepackage{rotating}
\usepackage{booktabs}
\usepackage{float}
\usepackage{color}
\usepackage{ulem}

\newcommand{\snu}{\affil{Astronomy Program, Department of Physics and Astronomy, Seoul National University, 1 Gwanak-ro, Gwanak-gu, Seoul 08826, Republic of Korea}}
\newcommand{\yonsei}{\affil{Department of Astronomy, Yonsei University, 50 Yonsei-ro, Seodaemun-gu, Seoul 03722, Korea} }

\shorttitle{star formation rates of AGN host galaxies}
\shortauthors{Kim et al.}

\begin{document}

\title{Determining star formation rates of active galactic nuclei host galaxies based on SED fitting with sub-mm data}

\correspondingauthor{Jong-Hak Woo}
\email{woo@astro.snu.ac.kr}

\author[0000-0002-2156-4994]{Changseok Kim}\snu
\author[0000-0002-8055-5465]{Jong-Hak Woo}\snu
\affil{SNU Astronomy Research Center, Seoul National University, 1 Gwanak-ro, Gwanak-gu, Seoul 08826, Republic of Korea}
\author{Yashashree Jadhav}\snu

\author{Aeree Chung}\yonsei
\author{Junhyun Baek}\yonsei
\author{Jeong Ae Lee}\snu
\author{Jaejin Shin}\snu
\affil{Department of Astronomy and Atmospheric Sciences, Kyungpook National University, Daegu 41566, Republic of Korea}
\author{Ho Seong Hwang}\snu
\affil{SNU Astronomy Research Center, Seoul National University, 1 Gwanak-ro, Gwanak-gu, Seoul 08826, Republic of Korea}
\author{Rongxin Luo}\snu
\author{Donghoon Son}\snu
\author{HyunGi Kim}\snu
\author{Hyuk Woo}\snu

\begin{abstract}

We present the star formation rate (SFR) measurements based on the spectral energy distribution (SED) analysis with new sub-mm fluxes combined with archival multi-wavelength data for a sample of 52 AGN host galaxies at z $<0.2$. We carried out sub-mm observations using the SCUBA-2 camera at the James Clerk Maxwell Telescope, and obtained flux or an upper limit at 450 and 850 $\mu$m for each target. 
By experimenting the effect of the AGN dust component in the SED fit, we find that dust luminosity can be overestimated if AGN contribution is ignored. While the SFR based on 4000\AA\ break shows a significant offset compared to dust luminosity based SFR, 
the SFR obtained by the artificial neural network \citep{Ellison16a} generally shows consistency albeit with a large scatter. We find that SFR correlates with AGN outflow strength manifested by the [\OIII] $\lambda 5007$ emission line, and that AGNs with higher Eddington ratios and stronger outflows are in general hosted by galaxies with higher SFR, which is consistent with the correlation reported by \cite{Woo:2020}. This suggests no instantaneous quenching of star formation due to AGN feedback. 
 \end{abstract}

\keywords{active galactic nuclei, star formation}

\section{Introduction} \label{sec:intro}

The coevolution scenario of supermassive black holes (SMBHs) and host galaxies assumes that active galactic nuclei (AGNs) play an important role in galaxy evolution by quenching or suppressing star formation \citep[negative feedback, e.g.,][]{Silk98, Fabian12, Shimizu+2015}. In contrast, there are also evidences of positive feedback such that AGN-driven outflows may trigger star formation by compressing interstellar medium (ISM) \cite[e.g.,][]{Cresci15, Maiolino17, Shin19}. While cosmological models routinely include the AGN feedback in various ways, i.e., radiative or mechanical feedback \citep[e.g.,][]{Shankar+13, Choi15, Croton+16} and numerous observational studies attempt to understand how AGN regulates star formation, especially by investigating gas outflows \citep[e.g.,][]{Cano-Diaz12, Sun14, Carniani16, Karouzos+16a,Karouzos16b, Woo:2017ApJ, Kang+18, Fluetsch19, Scholtz20, Woo:2020}, there is lack of direct observational evidences of quenching star formation by AGN feedback \citep{Harrison17}.

One of the observational limitations is that common star formation rate (SFR) indicators \citep[for a review, see][]{Kennicutt:1998ARA&A} suffer from contamination by AGN emission, leading to difficulties in measuring SFR. Therefore, calibrating SFR indicators for AGN host galaxies is a key to probe AGN feedback \citep[e.g.,][]{Woo12, Matsuoka:2015ApJ, Ellison16a, Shimizu17, Zhuang19O2, Riffel21}. Among various indicators, far-Infrared (FIR) emission reradiated by dust in the ISM is a good tracer since FIR contribution from Rayleigh-Jeans tail of hot dusty torus heated by AGN \citep{Netzer07} is negligible. A number of studies on AGN feedback adopted this FIR or total IR luminosity of dust emission as a primary SFR tracer \citep[e.g.,][]{Rosario12, Chen13, Matsuoka:2015ApJ, Shimizu+2015}.

In general, IR luminosity reradiated by dust in the ISM can be determined by SED modeling, using well-defined dust emission models \citep[e.g.,][]{Chary01,Draine&Li07, Draine14}. In the case of AGN host galaxies, however, there is a mid-Infrared (MIR) excess from hot dusty torus heated by AGN. Thus, decomposition of AGN-heated dust component is necessary to reliably measure the IR emission from interstellar dust, which peaks at $\sim$100 $\mu$m, and to determine star formation rate \citep[e.g.,][]{Fritz:2006MNRAS, Mullaney:2011MNRAS, CalistroRivera16, Leja18, Boquien19, Abdurro'uf21}.

In a series of papers, we investigated demography of ionized gas outflows traced by velocity offset and dispersion of [\OIII] $\lambda 5007 $ line, using a large sample of local type 1 and type 2 AGNs at low redshift, i.e., z$<$ 0.3 \citep{Woo+16, Rakshit18}, reporting that ionized gas outflows are prevalent both in type 1 and type 2 AGNs and outflow strength increases with AGN luminosity or Eddington ratio. By comparing these kinematic measurements with SFR estimates, which were based on the IR luminosity estimates
from Artificial Neural Network (ANN) analysis by \citet{Ellison16a} or FIR flux measurements \citep{Matsuoka:2015ApJ}, we found that AGNs with strong outflows (or high Eddington ratios) have higher or similar specific SFRs (sSFR) compared to main-sequence star-forming galaxies (SFGs), while AGNs with no/weak outflows (or low Eddington ratios) have lower sSFRs than SFGs \citep{Woo:2017ApJ, Woo:2020}. This result suggests
that star formation is not instantaneously suppressed when SMBH is still active.  Although we found a positive correlation between sSFR and AGN activity using a large sample of low-z AGNs, the systematic uncertainty of ANN-based SFR is one of the limitations of our results. Albeit a relatively small sample size, more detailed investigation along with SFR measurements based on SED fitting with multiwavelength data may provide necessary constrains to confirm the trends. 

In this paper, we present new sub-mm observations and flux measurements of 52 local AGNs, and their SFR estimates based on dust IR luminosity from SED fitting analysis to overcome the main limitation of our previous studies. Then, we explore the correlations between AGN activity (i.e., outflow strength or Eddington ratio) and star formation activity with the newly measured SFR based on IR luminosity, to investigate implications for AGN feedback scenarios. In Section \ref{sec:sam}, we describe the sample selection, sub-mm data reduction. In Section \ref{sec:SED}, we describe SED modeling analysis, and in Section \ref{sec:results} we present main results. Summary and conclusions follow in Section \ref{sec:con}.

\begin{figure}[h]
\includegraphics[width=\columnwidth]{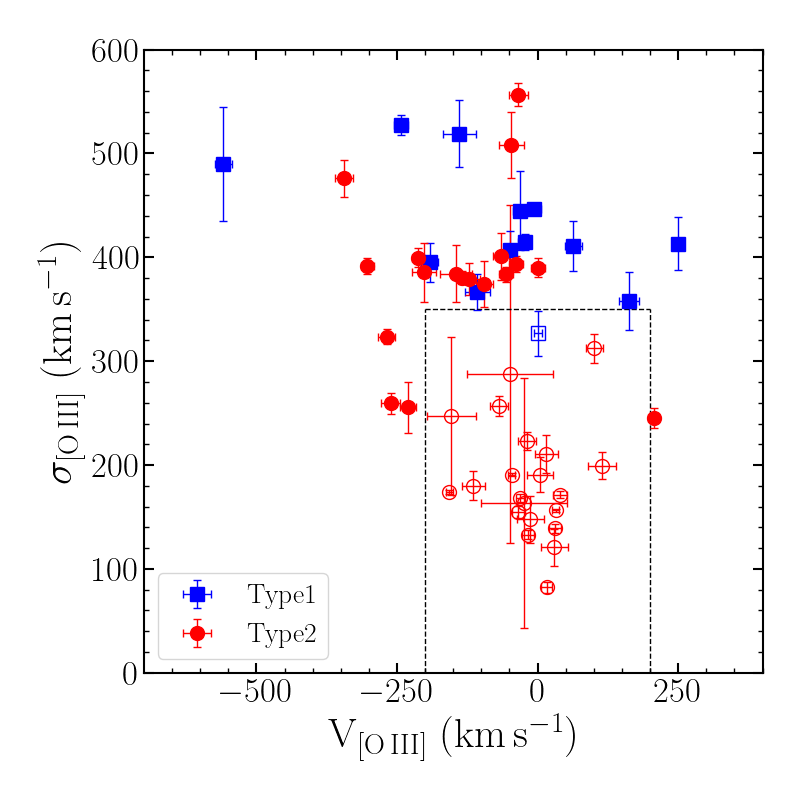}
\caption{Outflow properties of the type 1 (blue squares) and type 2 AGNs (red circles). The velocity (V$_{\rm{[OIII]}}$) and velocity dispersion ($\sigma_{\rm{[OIII]}}$) of the [\OIII] emission line were obtained from \citet{Woo+16}. We selected AGNs with strong outflows, i.e., [\OIII]\ velocity dispersion $>$ 350 \kms\ or [\OIII]\ velocity offset $>$ 200 \kms\ (filled symbols). AGNs without strong outflows (open symbol within the enclosed box) were also selected for comparison. 
}
\label{o3dispo3vel}
\end{figure}

    \begin{figure*}[h]
    \centering
    \scalebox{1}{
    \includegraphics[width=0.99\textwidth]{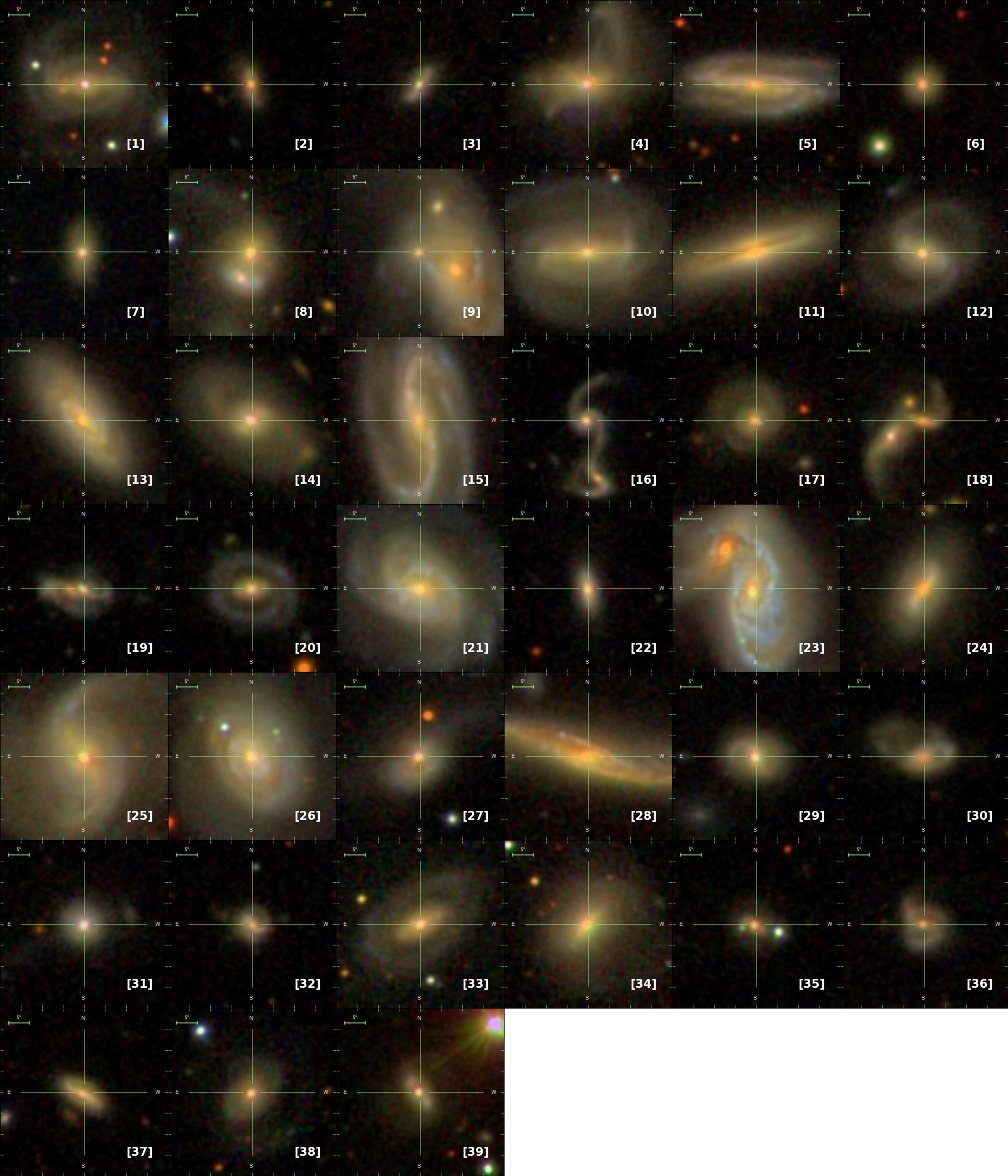}}
    \caption{Optical $gri$-composite images of type 2 AGNs from SDSS DR12. The FOV of each each panel is 40$\arcsec \times 40\arcsec$.}
    \label{SDSSt2}
    
    \end{figure*}
    
     \begin{figure*}
    \centering
    \scalebox{1}{
    \includegraphics[width=0.99\textwidth]{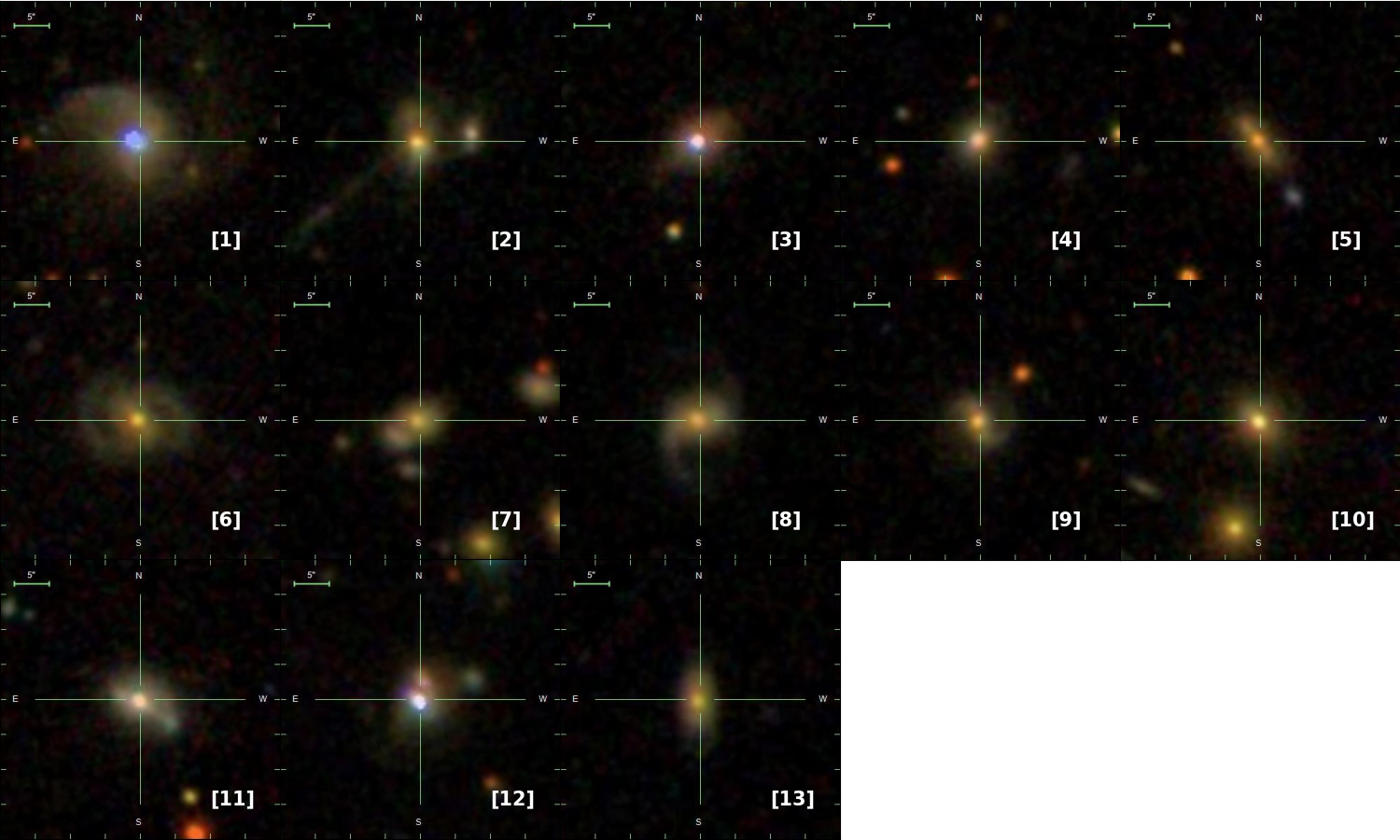}}
    \caption{Optical $gri$-composite images of type 1 AGNs from SDSS DR12. The FOV of each each panel is 40$\arcsec \times 40\arcsec$.}
    \label{SDSSt1}
    
    \end{figure*}

\section{sub-mm observations}\label{sec:sam}

\begin{figure*}[ht]

\centering
\includegraphics[height=4.5cm, width=0.99\textwidth]{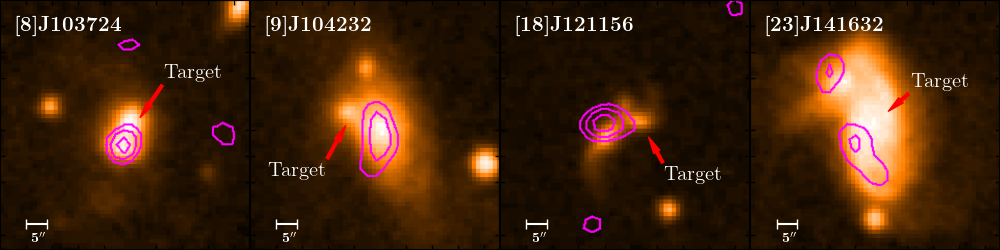}

\caption{Optical DSS images of interacting galaxies, overlaid with sub-mm contours, which indicate 2, 3, and 4$\sigma$ at 450 $\mu$m. Only one target, [23]J141632 shows two clumps with 3$\sigma$ detections. The size of each panel is $1'\times1'$.}
\label{Interactingeg}
\end{figure*}

In order to determine the total IR luminosity and SFR of AGN host galaxies, it is important to cover the large spectral range in the SED. FIR is a crucial regime to be included in the SED analysis since the black body radiation from dust peaks at FIR. Particularly for AGNs, FIR and sub-mm observations are of importance since MIR is contaminated by the contribution from hot dusty torus heated by AGN. 
Sub-mm flux can also play a key role in determining the total IR luminosity when FIR observation is not available. In this section we describe sample selection, sub-mm observations, and data reduction.

\subsection{Sample selection}

We selected a sample of 39 type 2 and 13 type 1 AGNs for sub-mm observations and SED analysis. 
First, we started with type 2 AGNs, using  the parent sample of 39,000 type 2 AGNs, which were identified based on the optical emission line ratios in our previous studies of AGN outflows based on the SDSS data \citep[see for details][]{Woo+16}.

For the initial observations in 2015, we selected a pilot sample of 18 type 2 AGNs at z $<$ 0.1, for which multi-wavelength data including FIR detections (i.e., AKARI or Herschel, see Section \ref{subsec:mlambda}) were available in the archive, without
strong constraints on outflow characteristics. 

After these pilot observations, we expanded the sample based on AGN luminosity and ionized gas outflow velocities, as we started optical IFU follow-up observations for strong outflow AGNs \citep{Karouzos+16a}.
We searched for strong outflow AGNs with [\OIII]\ velocity offset $>$ 200 \kms\ or [\OIII]\ velocity dispersion $>$ 350 \kms\ (i.e., [\OIII]\ FHWM $\gtrsim$1000 \kms), and initially obtained 116 objects from 23,000 type 2 AGNs at z $<$ 0.1 \citep{Bae+14}. Among these objects, we selected 15 type 2 AGNs as a secondary set for the JCMT observations in 2018. We also selected 6 weak outflow AGNs as a comparison sample, which do not show strong outflow signatures in the SDSS spectra, albeit with relatively high [\OIII] velocity \citep[see open symbols in Figure \ref{o3dispo3vel};][]{Luo+19}. Thus, by comparing SFR of type 2 AGNs with/without strong outflow features, we will be able to study how outflows (or luminosities of AGN) are related with host galaxy star formation. 
Note that all of these 21 objects were studied with spatially-resolved optical spectroscopy using either the Gemini/GMOS \citep{Karouzos+16a, Karouzos16b, Kang+18, Luo+19} or the VLT VIMOS \citep{Bae+17}. Combining with the pilot sample, we selected a total of 39 type 2 AGNs for sub-mm observations and SED analysis.

Second, we expanded the sample by including type 1 AGNs. Using the type 1 AGN catalog by \citet{Rakshit18}, we selected 13 targets at $0.08<z<0.2$ for JCMT/SCUBA-2 observations.
We applied similar selection criteria for choosing luminous AGNs with strong outflows, i.e.,  [\OIII]\ luminosity L$_{\rm{[OIII]}} > 10^{42}$ \ergs\ and [\OIII]\ velocity dispersion $\gtrsim$ 400 \kms. 
Note that type 1 and type 2 AGNs are complementary to each other since host galaxy properties, i.e., stellar mass, can be easily determined for type 2, while black hole mass and Eddington ratios are more securely measured for type 1 AGNs. Also, type 1 AGNs are on average more luminous and at higher z than type 2 AGNs.

In summary, we selected a sample of 52 AGNs, covering a large dynamic range of AGN luminosity and outflow strength. While the sample is not uniquely defined by any specific selection function as we selected subsamples somewhat differently, there is no strong selection bias. Thus, this sample with a broad range of AGN strength is useful for probing the correlation between AGN and star formation activity. We present optical images of the sample in Figure \ref{SDSSt2} and \ref{SDSSt1}, and the properties of each target in Table \ref{tablet2_fluxLIR} and \ref{tablet1_fluxLIR}.

\begin{table*}[h]
\centering
\caption{Sub-mm flux and IR luminosity of type 2 AGNs}
\label{tablet2_fluxLIR}
\begin{tabular}{ccccccccccc}
\hline
 & SDSS ID             & z & Morphology & 450 $\mu$m  & 450 $\mu$m$_{err}$   & 850 $\mu$m & 850 $\mu$m$_{err}$  & obs. &  T$_{int}$ &  log(L$_{IR})$  \\
       &                     &          &            & [mJy]         & [mJy]               & [mJy]         & [mJy] &  & [sec] & [erg/s]  \\
 {[}1{]}  & {[}2{]}  & {[}3{]}  & {[}4{]}     & {[}5{]}           & {[}6{]}      & {[}7{]}          & {[}8{]}     & {[}9{]}          & {[}10{]}& {[}11{]} \\
       \midrule
1	   & J075328.31+142141.0 & 0.05		& S 		 &	136$^{\star}$		&45			&	6$^{\star}$	& 2      & 18B	        &	3652	& 44.47$\pm$0.02    \\
2      & J082714.96+263618.2 & 0.09     & U   		 & 193$^{\star}$  		& 64	    & 9$^{\star}$ 	& 3      & 18B          & 1831      & 44.54$\pm$0.16    \\
3      & J083132.28+160143.3 & 0.07     & U          & 101$^{\star}$  		& 34	    & 5$^{\star}$  	& 2      & 18B          & 7347      & 43.68$\pm$0.10    \\
4      & J084344.98+354942.0 & 0.05     & U          & -                  	& - 	   	& -             & -      & -      	    &   -       & 44.21$\pm$0.02    \\
5      & J085547.66+004739.4 & 0.04     & S          & 306$^{\star}$   		& 102	   	& 13            & 2      & 18A          & 8725      & 44.49$\pm$0.02    \\
6      & J091807.52+343946.0 & 0.10     & U          & 891$^{\star}$ 		& 297	   	& 6$^{\star}$   & 2      & 18A          & 17562     & 44.92$\pm$0.02    \\
7      & J101936.79+193313.4 & 0.06     & U          & 218$^{\star}$  		& 73	   	& 6$^{\star}$   & 2      & 18B          & 7338      & 43.73$\pm$0.12    \\
8      & J103723.62+021845.5 & 0.04     & M          & 88              		& 19	    & 9$^{\star}$   & 3      & 15A          & 1081      & 44.70$\pm$0.02    \\
9      & J104232.05+050241.9 & 0.03     & M          & 61              		& 14	    & 15            & 3      & 15A          & 1579      & 44.03$\pm$0.02    \\
10     & J105833.33+461604.8 & 0.04     & S          & 81$^{\star}$  		& 27	    & 4$^{\star}$  	& 1      & 18B          & 7364      & 43.61$\pm$0.02    \\
11     & J110037.22+112455.1 & 0.03     & S          & -                  	& -         & -             & -      & -            &    -      & 43.63$\pm$0.02    \\
12     & J110630.64+063333.9 & 0.04	    & S 		 & 505$^{\star}$		& 168		&	5$^{\star}$	& 2	     & 18A	        & 17570 	& 44.33$\pm$0.02    \\
13     & J111406.30+554239.1 & 0.03     & S          & 177             		& 25		& 23            & 4      & 15A          & 947       & 44.38$\pm$0.02    \\
14     & J113549.07+565708.2 & 0.05     & S          & 174 					& 25	    & 6$^{\star}$  	& 2      & 15A          & 824       & 44.95$\pm$0.03    \\
15     & J113606.63+621456.9 & 0.03     & S          & 89              		& 27	    & 19           	& 5      & 15A          & 885       & 44.59$\pm$0.02    \\
16	   & J114719.93+075243.0 & 0.08	    & U 		 & 84$^{\star}$			& 28		& 4$^{\star}$	& 1	     & 18B	        & 7358	    & 44.82$\pm$0.06	\\
17     & J115657.88+550821.5 & 0.08     & S          & -                  	& - 	    & -             & -      & -            &  -        & 44.16$\pm$0.05    \\
18     & J121155.63+372113.4 & 0.08     & M          & 93              		& 20	    & 15            & 3      & 15A          & 1369      & 44.05$\pm$0.16    \\
19     & J125642.71+350729.9 & 0.05     & U          & 79$^{\star}$         & 26 	  	& 14$^{\star}$  & 5      & 15A          & 635       & 45.03$\pm$0.02    \\
20     & J131153.80+053138.3 & 0.09     & S          & 105$^{\star}$   		& 35	    & 5$^{\star}$  	& 2      & 18B          & 7346      & 44.03$\pm$0.18    \\
21     & J132948.19+310748.5 & 0.02     & S          & 119                  & 27 	 	& 17            & 3      & 15A          & 1562      & 44.12$\pm$0.02    \\
22     & J140452.65+532332.1 & 0.08     & U          & 1119$^{\star}$ 		& 373	    & 6$^{\star}$  	& 2      & 18A          & 13278     & 44.68$\pm$0.08    \\
23	   & J141631.74+393521.2 & 0.03		& U			 & 402					& 57		& 28        	& 6      & 15A	        & 336		& 44.44$\pm$0.02	\\
24     & J142859.53+605000.5 & 0.05     & S          & 244              	& 58	    & 23            & 5      & 15A          & 580       & 44.85$\pm$0.02    \\
25     & J143545.74+244332.8 & 0.04     & S          & 179            		& 43	 	& 13$^{\star}$  & 4      & 15A          & 889       & 44.57$\pm$0.02    \\
26     & J145835.98+445300.9 & 0.04     & S          & 215              	& 55	    & 27            & 6      & 15A          & 396       & 44.91$\pm$0.02    \\
27     & J152549.54+052248.7 & 0.05     & E          & 173              	& 44	    & 19$^{\star}$  & 6      & 15A          & 332       & 45.16$\pm$0.05    \\
28     & J160507.88+174527.6 & 0.03     & S          & 178             		& 25	    & 21$^{\star}$  & 7      & 15A          & 638       & 44.38$\pm$0.03    \\
29     & J160652.16+275539.1 & 0.05     & U          & 900$^{\star}$ 		& 300	    & 6$^{\star}$ 	& 2      & 18A          & 13142     & 44.21$\pm$0.11    \\
30     & J161534.13+210019.7 & 0.09     & U          & 135             		& 29		& 22$^{\star}$  & 7      & 15A          & 521       & 45.27$\pm$0.06    \\
31     & J162232.68+395650.2 & 0.06     & U          & 883$^{\star}$ 		& 294		& 6$^{\star}$   & 2		 & 18A          & 13162     & 44.55$\pm$0.02    \\
32     & J172037.94+294112.4 & 0.10     & S          & 392$^{\star}$  		& 131		& 5$^{\star}$  	& 2      & 18A          & 13151     & 44.59$\pm$0.07    \\
33	   & J203907.05+003316.3 & 0.05     & S 		 & 195$^{\star}$		& 65		& 9$^{\star}$	& 3	     & 18B      	& 1816		& 44.57$\pm$0.02	\\
34     & J205536.51-003811.7 & 0.05     & E          & 310$^{\star}$  		& 103		& 9$^{\star}$ 	& 3      & 18B          & 1828      & 43.17$\pm$0.06    \\
35     & J210506.94+094118.8 & 0.10     & M          & 114$^{\star}$   		& 38		& 5$^{\star}$  	& 2      & 18B          & 7324      & 44.64$\pm$0.07    \\
36     & J211307.21+005108.4 & 0.07     & U          & 99$^{\star}$  		& 33		& 5$^{\star}$ 	& 2      & 18B          & 7366      & 43.98$\pm$0.19    \\
37     & J211333.79-000950.4 & 0.07     & U          & 135$^{\star}$  		& 45		& 6$^{\star}$ 	& 2      & 18B          & 3650      & 44.52$\pm$0.07    \\
38     & J213333.31-071249.2 & 0.09     & S          & 237$^{\star}$   		& 79		& 4$^{\star}$  	& 1      & 18A          & 13173     & 45.38$\pm$0.03    \\
39     & J214559.99+111325.9 & 0.09     & S          & 822$^{\star}$        & 274	    & 6  		    & 2      & 18A          & 13409     & 44.37$\pm$0.07    \\
\hline

\end{tabular}
\tablecomments{(1) target number. ; (2) SDSS name.; (3) redshift from SDSS DR 14.; (4) morphology based on SDSS and NASA Extragalactic Database. S, E, U, and M represent spiral, elliptical, uncertain, merging sources, respectively.; (5) JCMT-SCUBA 2 measurement at 450 $\mu$m. $^{\star}$ indicates upper limits.; (6) 450 $\mu$m errors.; (7) JCMT-SCUBA 2 measurement at 850 $\mu$m. $^{\star}$ indicates upper limits.; (8) 850 $\mu$m errors.; (9) Observation semester.; (10) Integration time of JCMT observation.; (11) dust IR luminosity based on the SED fitting.}
\end{table*}

\begin{table*}[]
\centering
\caption{Sub-mm flux and IR luminosity of type 1 AGNs}
\label{tablet1_fluxLIR}
\begin{tabular}{ccccccccccc}
\toprule
\hline
 & SDSS ID             & z & Morphology & 450 $\mu$m  & 450 $\mu$m$_{err}$   & 850 $\mu$m & 850 $\mu$m$_{err}$  & obs. &  T$_{int}$ &  log(L$_{IR})$  \\
       &                     &          &            & [mJy]         & [mJy]               & [mJy]         & [mJy] &  & [sec] & [erg/s]  \\
 {[}1{]}  & {[}2{]}  & {[}3{]}  & {[}4{]}     & {[}5{]}           & {[}6{]}      & {[}7{]}          & {[}8{]}     & {[}9{]}          & {[}10{]}& {[}11{]} \\
       \midrule
1      & J015950.23+002340.9 & 0.16  & U          & 469$^{\star}$  			& 156  & 9$^{\star}$ 			& 3     & 19A          & 1878       & 46.14$\pm$0.02       \\
2      & J020713.32-011223.1 & 0.17  & M          & 314$^{\star}$  			& 105  & 6$^{\star}$ 			& 2     & 19A          & 7517       & 45.29$\pm$0.05       \\
3      & J044428.77+122111.7 & 0.09  & U          & 461$^{\star}$  			& 154  & 10$^{\star}$ 			& 3     & 19A          & 1880       & 45.55$\pm$0.03       \\
4      & J073638.86+435316.5 & 0.11  & U          & 893$^{\star}$		    & 298  & 10$^{\star}$ 			& 3     & 19A          & 3784       & 45.41$\pm$0.12       \\
5      & J075244.63+434105.3 & 0.18  & U          & 446$^{\star}$ 			& 149  & 6$^{\star}$  			& 2     & 19A          & 7557       & 45.53$\pm$0.02       \\
6      & J084028.61+332052.2 & 0.17  & S          & 527$^{\star}$  			& 176  & 6$^{\star}$ 			& 2     & 19A          & 7568       & 44.53$\pm$0.02        \\
7      & J090403.72+074819.3 & 0.15  & M          & 370$^{\star}$  			& 123  & 6$^{\star}$  			& 2     & 19A          & 7562       & 45.15$\pm$0.07       \\
8      & J113320.56-033337.5 & 0.12  & S          & 319$^{\star}$  			& 106  & 6$^{\star}$  			& 2     & 19A          & 7570       & 44.87$\pm$0.08       \\
9      & J135617.79-023101.5 & 0.13  & S          & 329$^{\star}$  			& 110  & 6$^{\star}$ 			& 2     & 19A          & 6735       & 45.04$\pm$0.03       \\
10     & J140845.73+353218.4 & 0.17  & E          & 233$^{\star}$  			& 78   & 4$^{\star}$ 			& 1     & 19A          & 11376      & 44.63$\pm$0.23	    \\
11     & J142230.34+295224.2 & 0.11  & U          & 465$^{\star}$  			& 155  & 8$^{\star}$ 			& 3     & 19A          & 3783       & 45.38$\pm$0.02       \\
12     & J150913.79+175710.0 & 0.17  & M          & 344$^{\star}$  			& 115  & 7$^{\star}$ 			& 2     & 19A          & 5401       & 45.61$\pm$0.02       \\
13     & J225510.11-081234.2 & 0.15  & S          & 197$^{\star}$  			& 66   & 5$^{\star}$			& 2     & 19A          & 7511       & 45.43$\pm$0.05        \\
\hline
\end{tabular}
\tablecomments{(1) target number. ; (2) SDSS name.; (3) redshift from SDSS DR 14.; (4) morphology based on SDSS and NASA Extragalactic Database. S, E, U, and M represent spiral, elliptical, uncertain, merging sources, respectively.; (5) JCMT-SCUBA 2 measurement at 450 $\mu$m. $^{\star}$ indicates upper limits.; (6) 450 $\mu$m errors.; (7) JCMT-SCUBA 2 measurement at 850 $\mu$m. $^{\star}$ indicates upper limits.; (8) 850 $\mu$m errors.; (9) Observation semester.; (10) Integration time of JCMT observation.; (11) dust IR luminosity based on the SED fitting.}
\end{table*}

\subsection{SCUBA-2 observations}

We performed sub-mm observations at the James Clerk Maxwell Telescope (JCMT) with the Submillimeter Common-User Bolometer Array 2 (SCUBA-2) \citep{Holland:2013MNRAS} and obtained 450 and 850 $\mu$m flux or upper limit measurements (see Table \ref{tablet2_fluxLIR} and \ref{tablet1_fluxLIR}). In 2015, we observed 15 AGNs from the initial set of 18 type 2 AGNs (M15AI147; PI. Karouzos). For the other 3 targets, JCMT observations were not performed due to the limited telescope time. Nonetheless, we included them in our SED analysis since various IR flux measurements, e.g., Herschel/the Photoconductor Array Camera and Spectrometer (PACS) or the Spectral and Photometric Imaging Receiver (SPIRE), were available. For the 2nd set of type 2 AGNs, we observed 9 AGNs in 2018A (M18AP071; PI. Woo) and 12 AGNs in 2018B (M18BP070) out of the second set of 21 AGNs. Thus, the type 2 AGN sample is composed of 39 targets with our SCUBA-2 observations and 3 additional targets without new observations.

In 2019, we observed 13 type 1 AGNs (M19AP057; PI. Woo). Combining the type 1 and 2 AGNs, we obtained sub-mm observations at 450 and 850 $\mu$m for
49 targets except for 3 targets in the total sample of 52 objects. Given that our targets are unlikely to be very extended, we adopted the Daisy mapping mode which allows efficient mapping of small and compact sources with high sensitivity. The observations were performed under the Band 1 and 3 weather conditions, respectively in 2015 and 2018--2019. The on-source exposure time per each target ranged from $\sim$ 0.5 to $\sim$4 hours. 

\subsection{Data reduction}

The data were processed using the SCUBA-2 reduction pipeline \texttt{ORAC-DR} \citep{Jenness+15} in \texttt{Starlink} software \citep{Currie+14_Starlink}, with the recipe \texttt{REDUCE SCAN FAINT POINT SOURCES}. This recipe reduced scan data with a matched filter size as 7.9 \arcsec$\;$(450 $\mu$m) and 13\arcsec$\;$(850 $\mu$m), and estimated noises. After image processing, we cropped the images into a size of a 120\arcsec$\;$radius circle around each target by using \texttt{PICARD} package.

With these cropped images, we performed several steps of analysis to determine detections or upper limits. First, we compared target positions in optical images from the Digitized Sky Survey (DSS) at ESO with $3 \sigma$ sub-mm contours by using overlay images. If a $3 \sigma$ sub-mm contour is located on the target position in the DSS image, we classified it as a detection. Second, if there is a $>2\sigma$ clump at the DSS target position, we smoothed the image with a factor of three larger filter size to increase S/N. If the clump is higher than $3 \sigma$ of the smoothed data, we also classified it as detected. If there is no $> 3 \sigma$ clump at the target position even after smoothing, $3 \sigma$ value was adopted as a flux upper limit.

We measured the fluxes by reading the peak values of the images in the case of the point sources. However, 12 targets show extended structures in the images, so we performed a clump finding process for those targets by using \texttt{FellWalker} algorithm in \texttt{CUPID} package \citep{Berry+07, Berry+15}. This algorithm finds clumps in the given image by walking through pixels in every route heading the local maximums.
In Table \ref{tablet2_fluxLIR} and \ref{tablet1_fluxLIR}, we listed fluxes or upper limits at 450 and 850 $\mu$m, measured by SCUBA-2 observations. We detected total 14 targets at 450 $\mu$m, and 10 targets at 850 $\mu$m. All detected targets are type 2 AGNs, while there is no detection among type 1 AGNs at both 450 and 850 $\mu$m.

\subsection{Interacting Galaxies}

Among the detected targets, four type 2 AGNs (([8]J103724, [9]J104232, [18]J121156 and [23]J141632) showed disturbed morphology with multiple components in the optical image, which can be classified as interacting (merging) galaxies (Figure \ref{Interactingeg}). Because of the limited angular resolution in the images of FIR to sub-mm observations, it is hard to extract fluxes only from the component in which AGN is located. Therefore, we treat them as a single galaxy in the SED analysis.

While three of the interacting galaxies showed only one clump in the SCUBA-2 image, one galaxy [23]J141632 shows two 450 $\mu$m clumps located on each interacting pair (see Figure \ref{Interactingeg}). Thus, we adopt the sum of the fluxes from the two components as a total flux in SED analysis of this target. For consistency, we adopted the sum of stellar masses and D$_n$4000-based SFR of the two components for [18]J121156, and the stellar masses of the three components for [9]J104232 \citep{Barth08}, if the measurement for each component is available. In the case of ANN-based SFR, we found no measurements for these merging targets. Since ANN, D$_n$4000-based SFRs and other physical parameters are very uncertain due to the merging nature of the target, we exclude them in the following analysis. Note that with or without adding these merging galaxies, there is no significant change in the results.

\section{Spectral Energy Distribution analysis} \label{sec:SED}

\begin{figure*}[htb!]

\includegraphics[height=4.5cm, width=0.33\textwidth]{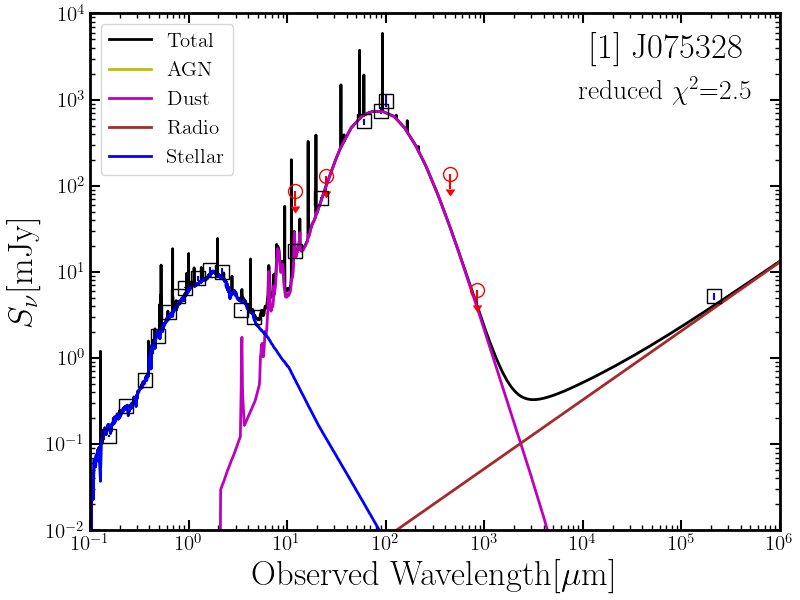}
\includegraphics[height=4.5cm, width=0.33\textwidth]{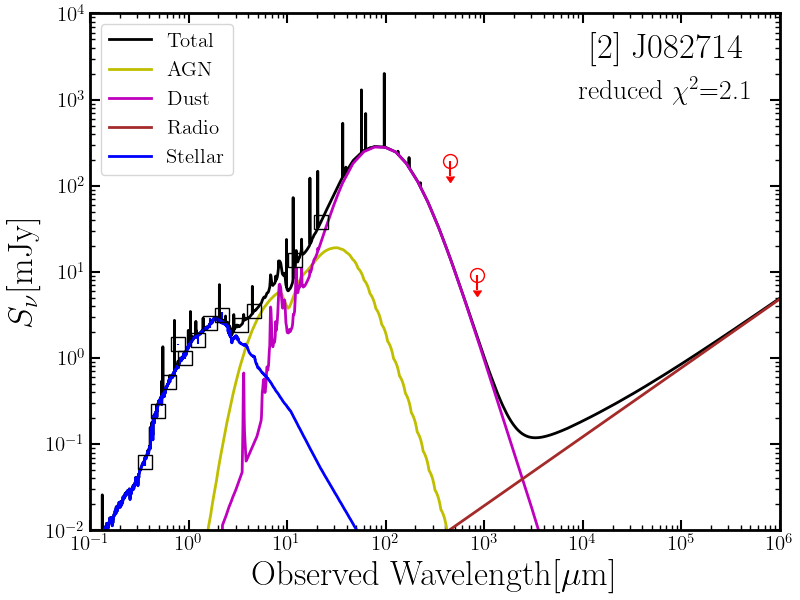}
\includegraphics[height=4.5cm, width=0.33\textwidth]{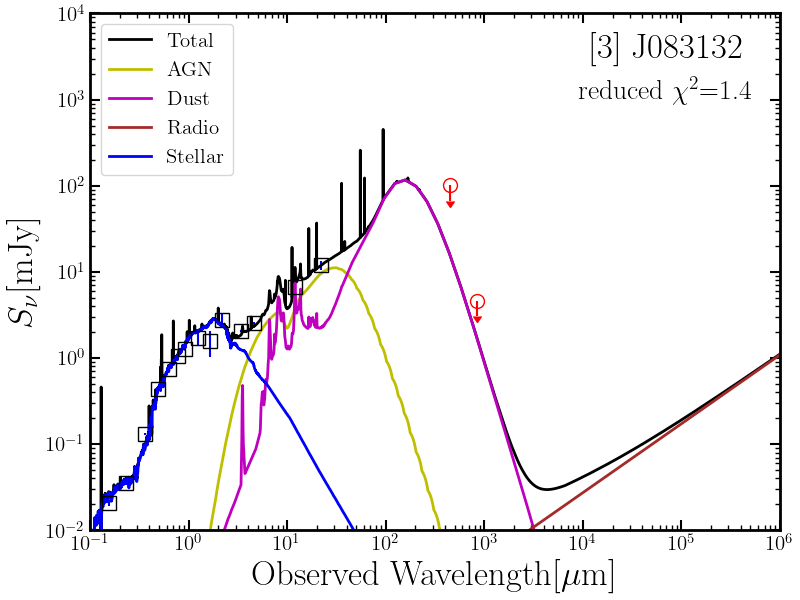}\\
\includegraphics[height=4.5cm, width=0.33\textwidth]{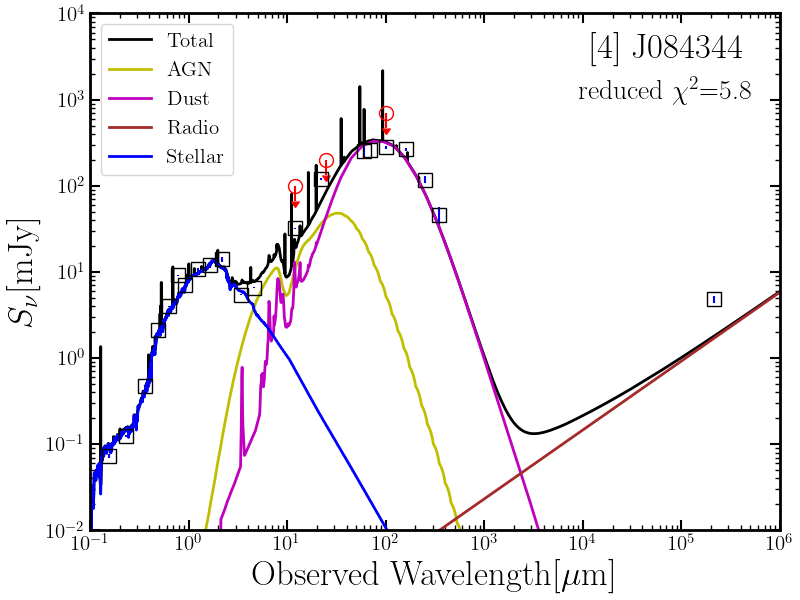}
\includegraphics[height=4.5cm, width=0.33\textwidth]{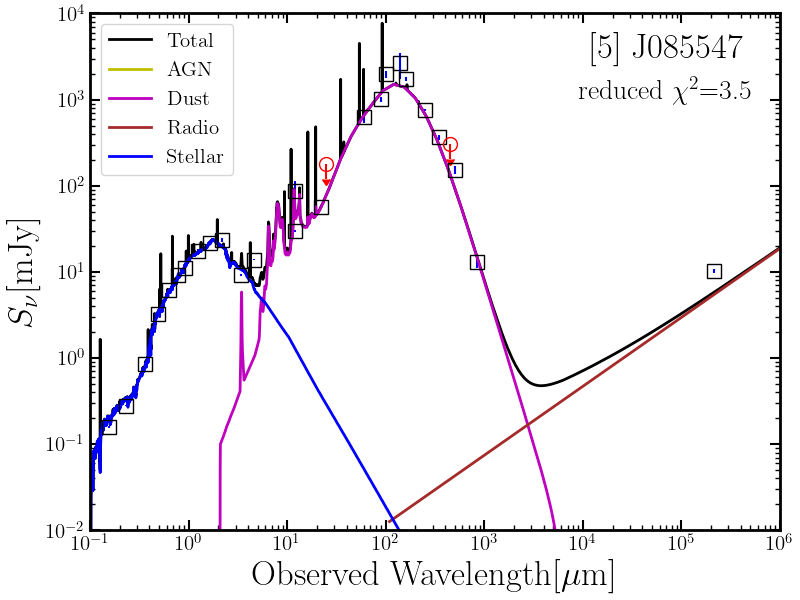}
\includegraphics[height=4.5cm, width=0.33\textwidth]{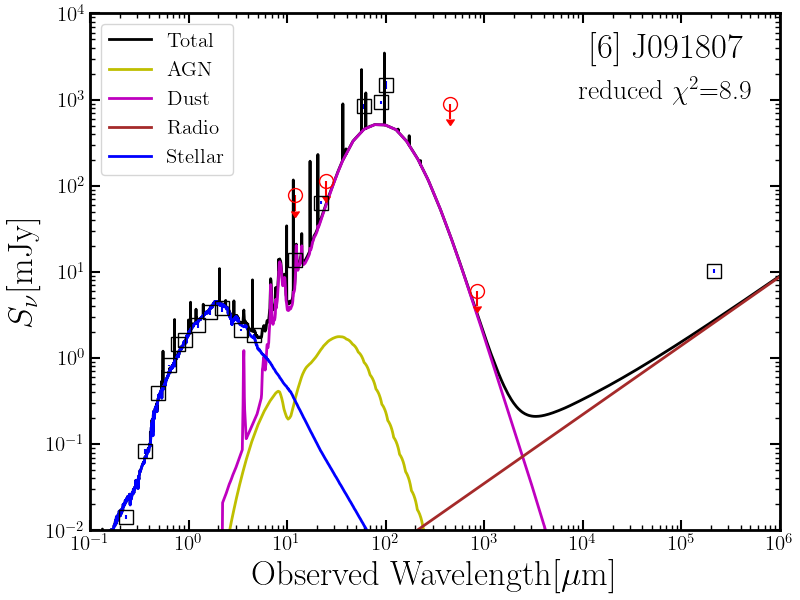}\\
\includegraphics[height=4.5cm, width=0.33\textwidth]{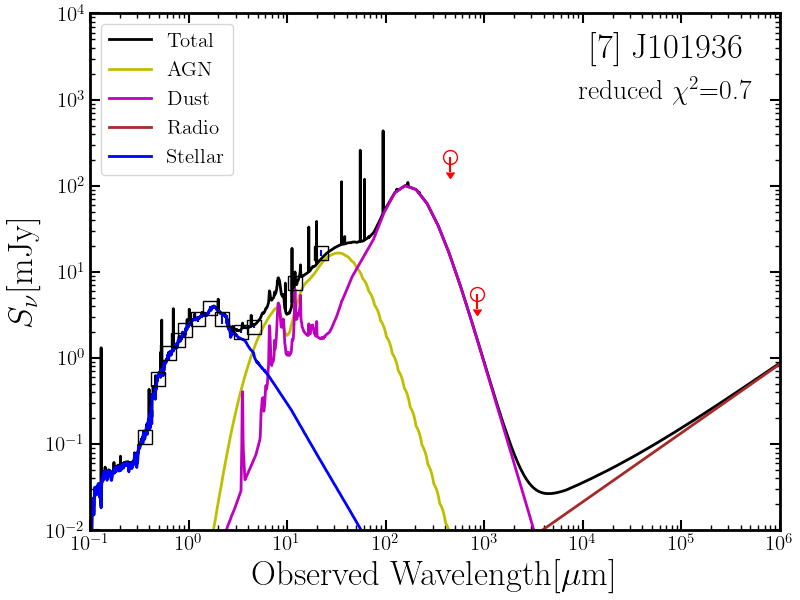}
\includegraphics[height=4.5cm, width=0.33\textwidth]{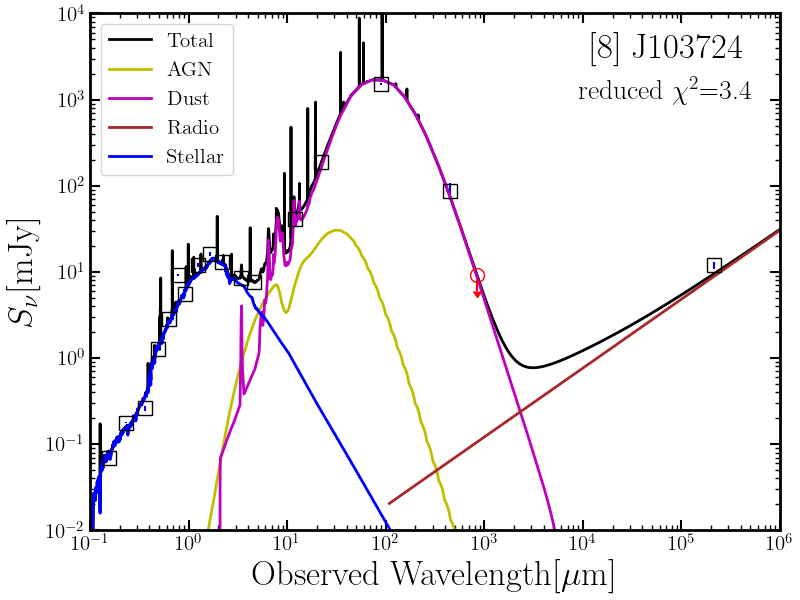}
\includegraphics[height=4.5cm, width=0.33\textwidth]{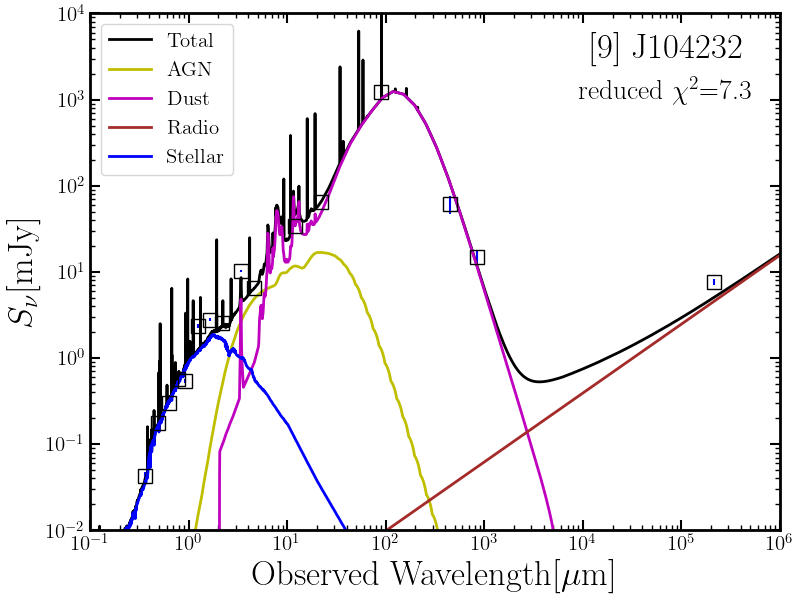}\\
\includegraphics[height=4.5cm, width=0.33\textwidth]{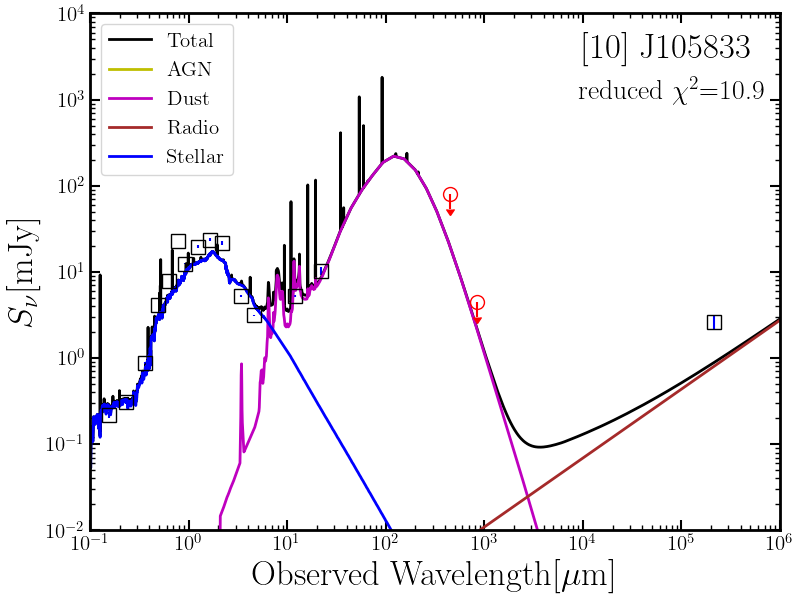}
\includegraphics[height=4.5cm, width=0.33\textwidth]{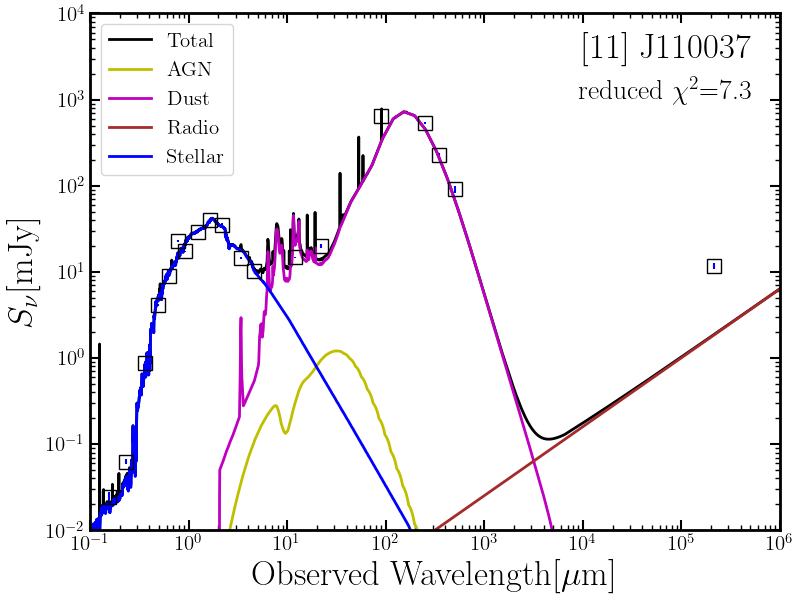}
\includegraphics[height=4.5cm, width=0.33\textwidth]{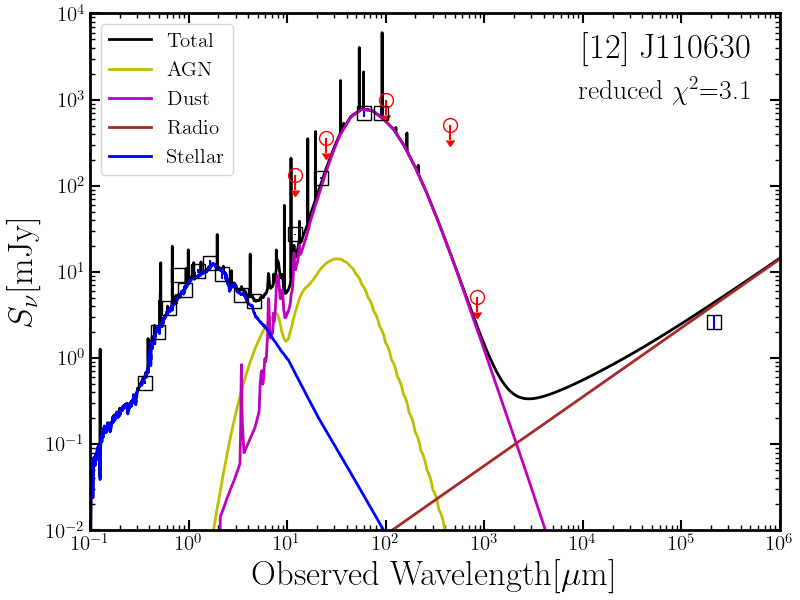}\\
\includegraphics[height=4.5cm, width=0.33\textwidth]{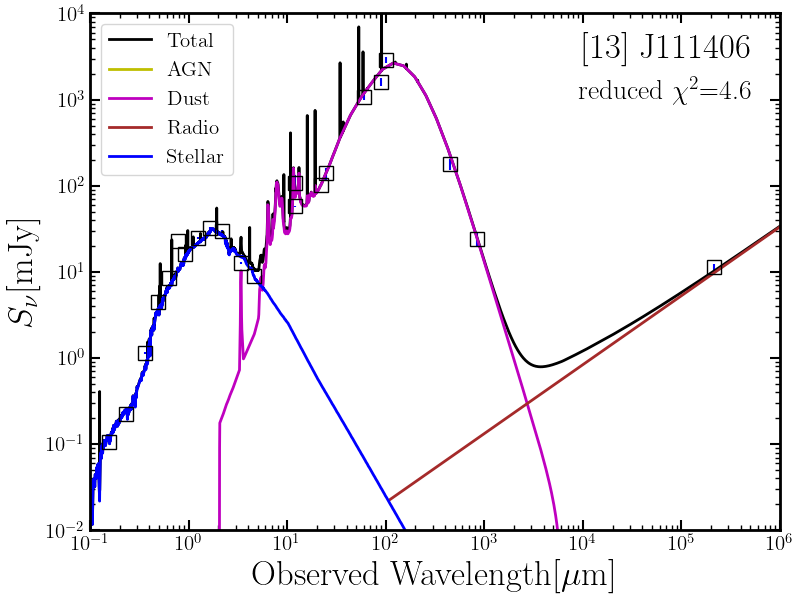}
\includegraphics[height=4.5cm, width=0.33\textwidth]{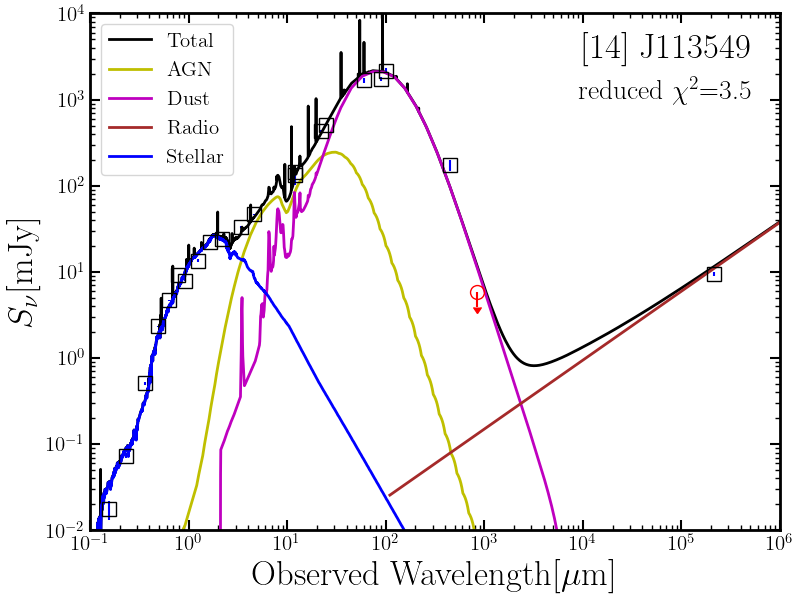}
\includegraphics[height=4.5cm, width=0.33\textwidth]{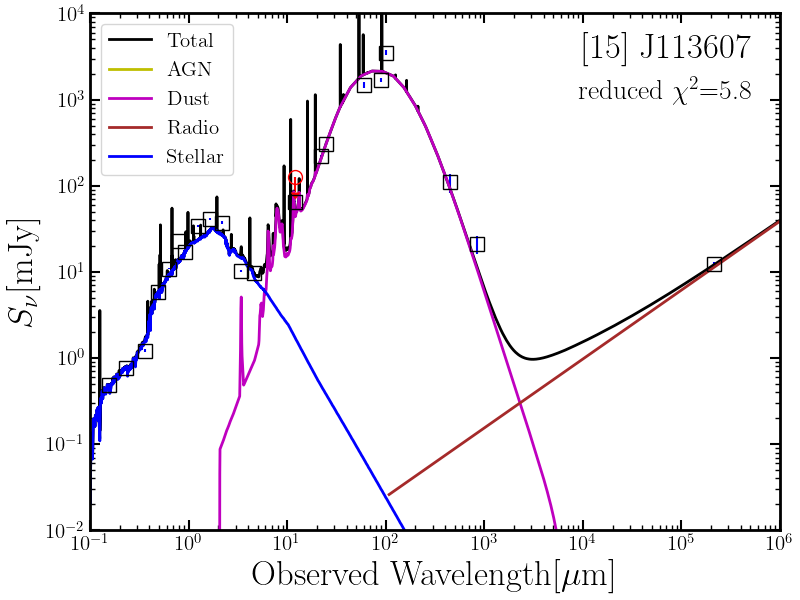}\\

\caption{The best-fit SED model (black line) of 39 type 2 AGNs. A dust component (magenta line) and an AGN component (green line) are also shown.}
\label{cigaleeg1}
\end{figure*}

\addtocounter{figure}{-1}
\begin{figure*}[htb!]
\includegraphics[height=4.5cm, width=0.33\textwidth]{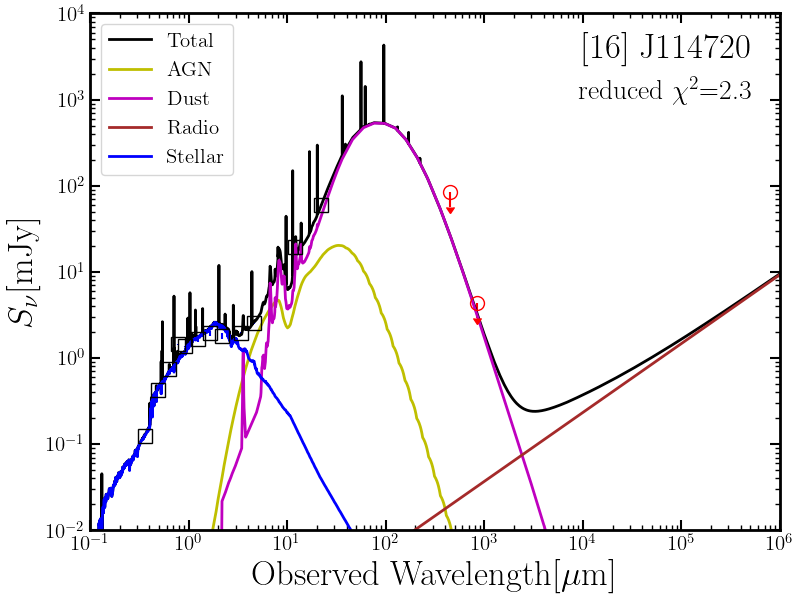}
\includegraphics[height=4.5cm, width=0.33\textwidth]{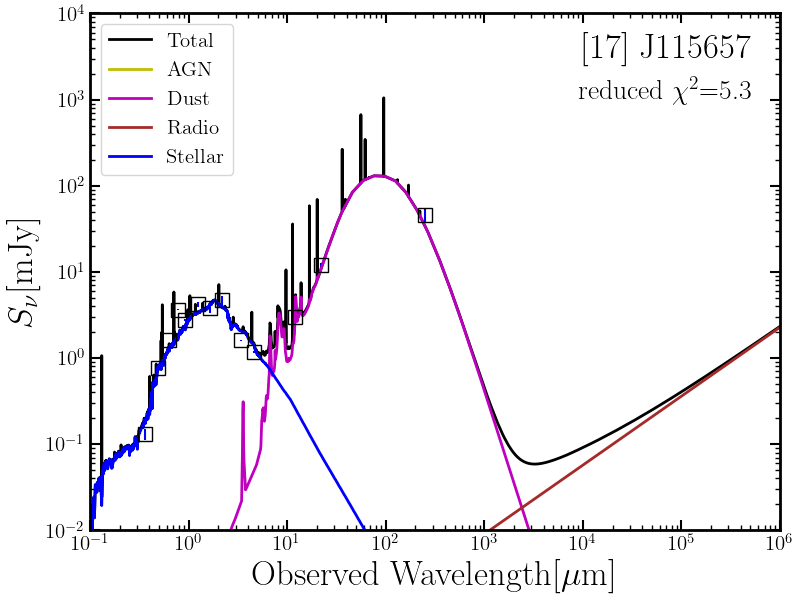}
\includegraphics[height=4.5cm, width=0.33\textwidth]{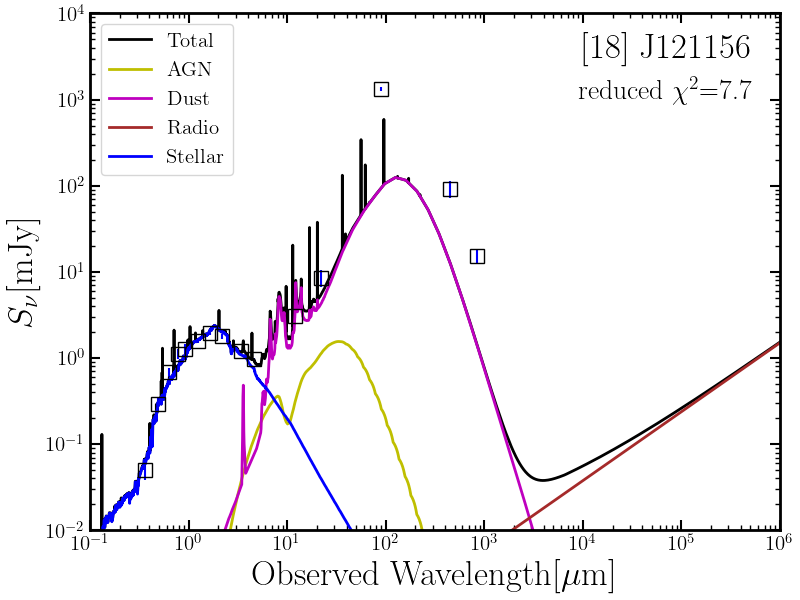}\\
\includegraphics[height=4.5cm, width=0.33\textwidth]{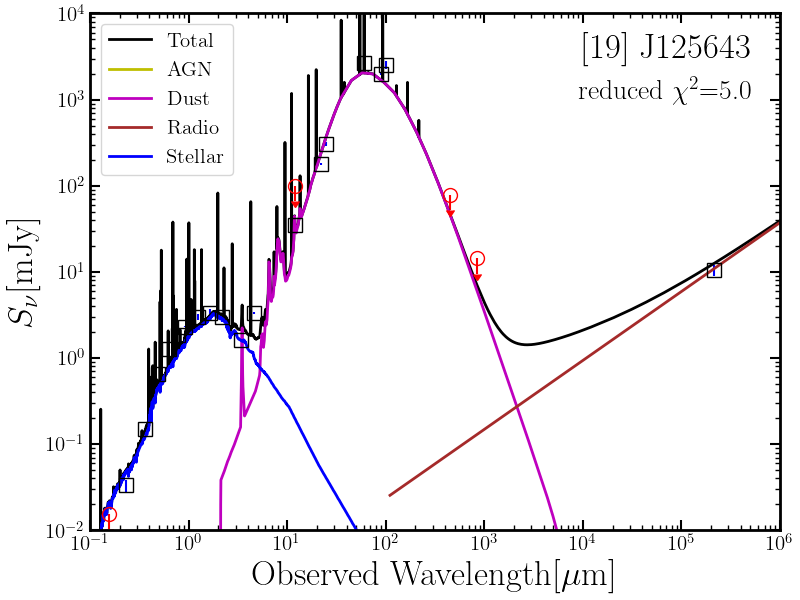}
\includegraphics[height=4.5cm, width=0.33\textwidth]{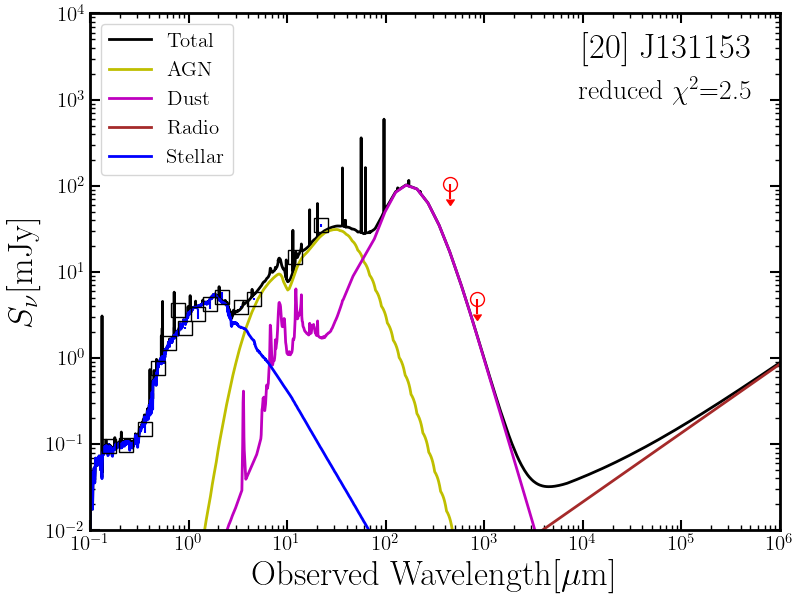}
\includegraphics[height=4.5cm, width=0.33\textwidth]{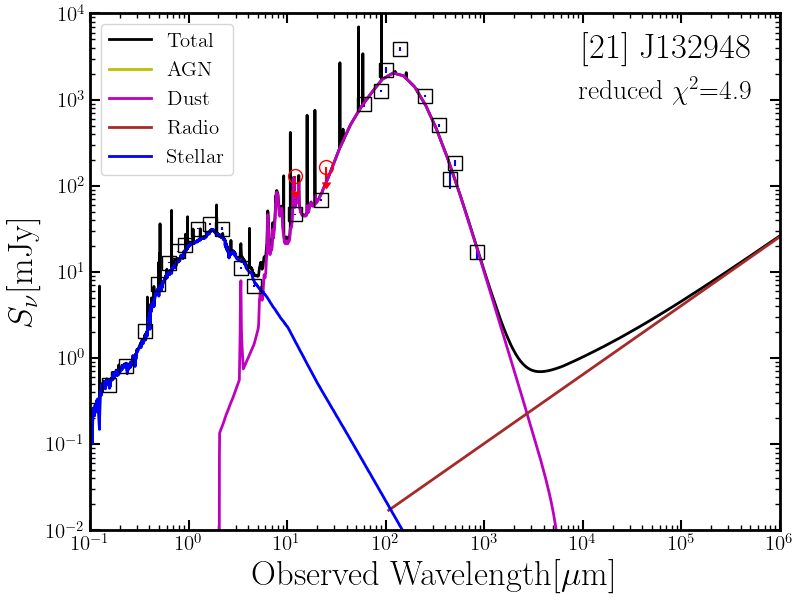}\\
\includegraphics[height=4.5cm, width=0.33\textwidth]{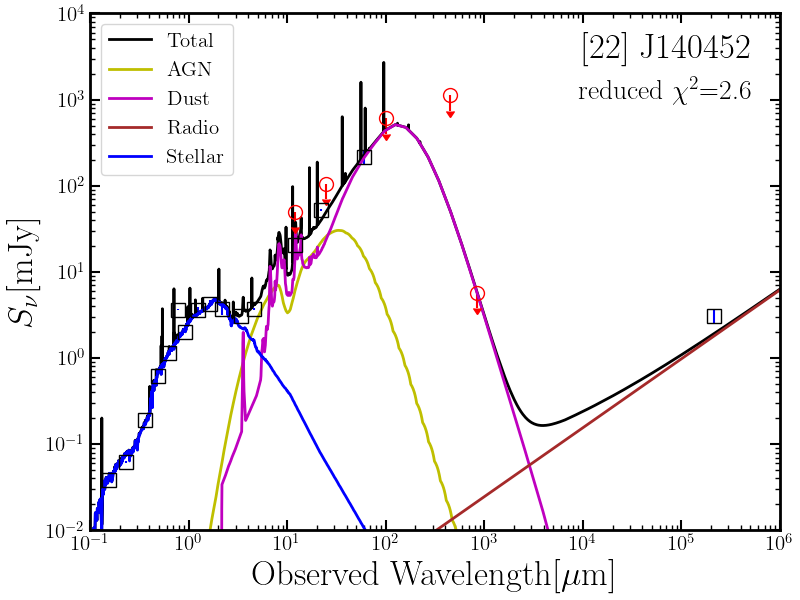}
\includegraphics[height=4.5cm, width=0.33\textwidth]{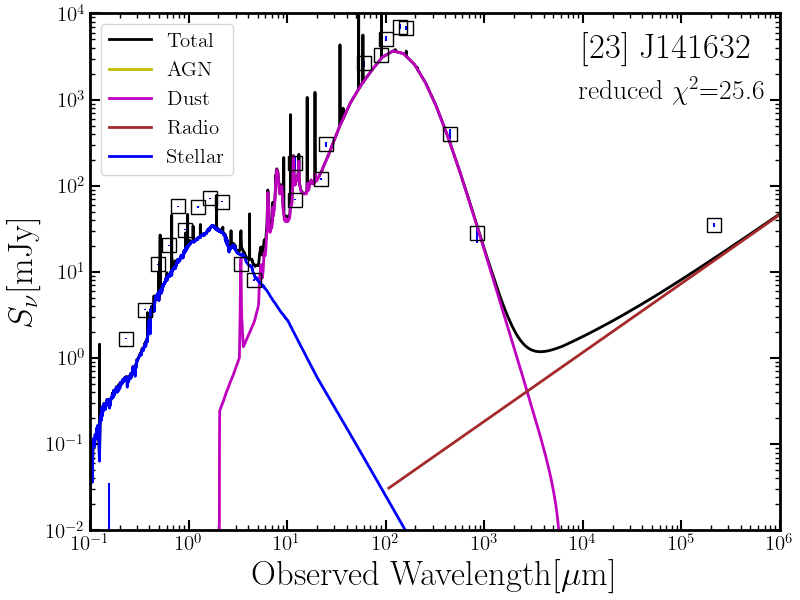}
\includegraphics[height=4.5cm, width=0.33\textwidth]{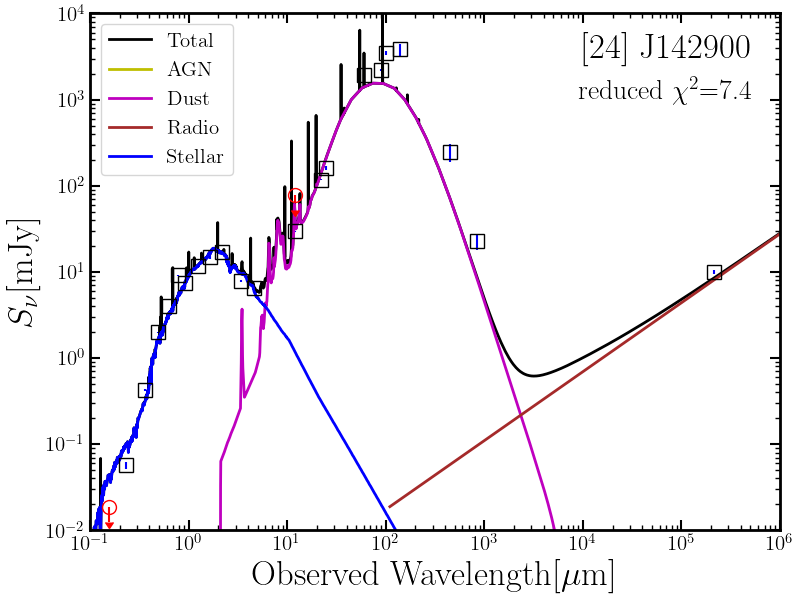}\\
\includegraphics[height=4.5cm, width=0.33\textwidth]{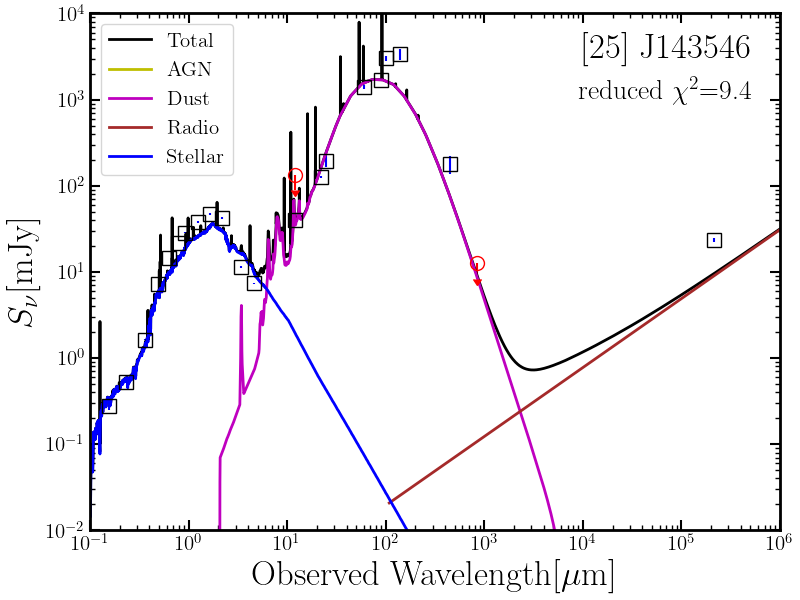}
\includegraphics[height=4.5cm, width=0.33\textwidth]{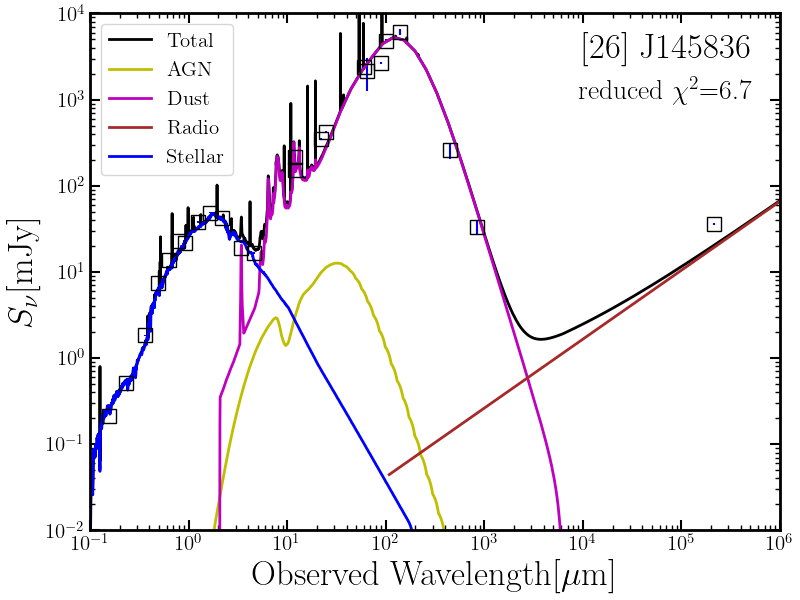}
\includegraphics[height=4.5cm, width=0.33\textwidth]{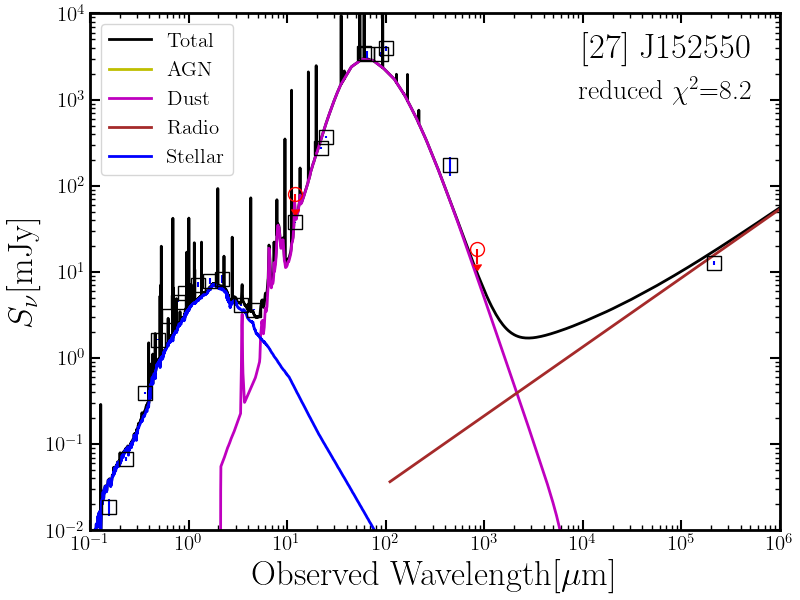}\\
\includegraphics[height=4.5cm, width=0.33\textwidth]{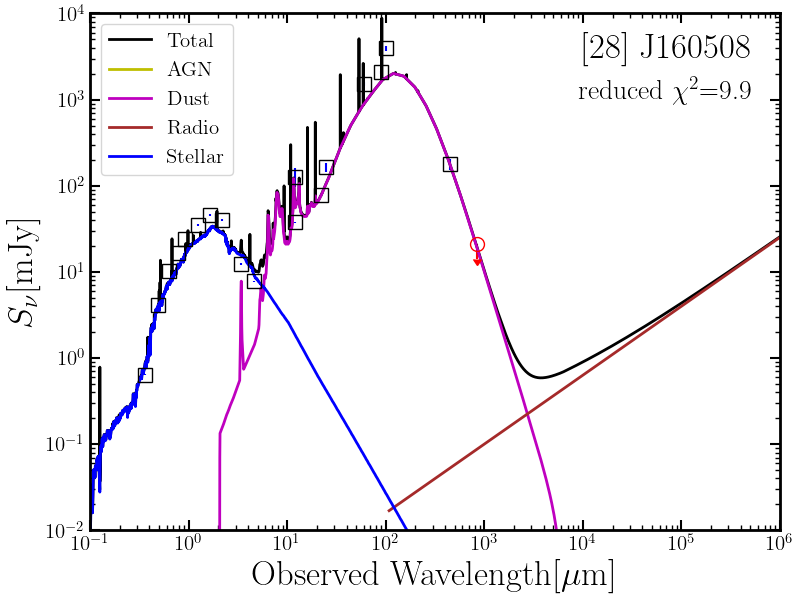}
\includegraphics[height=4.5cm, width=0.33\textwidth]{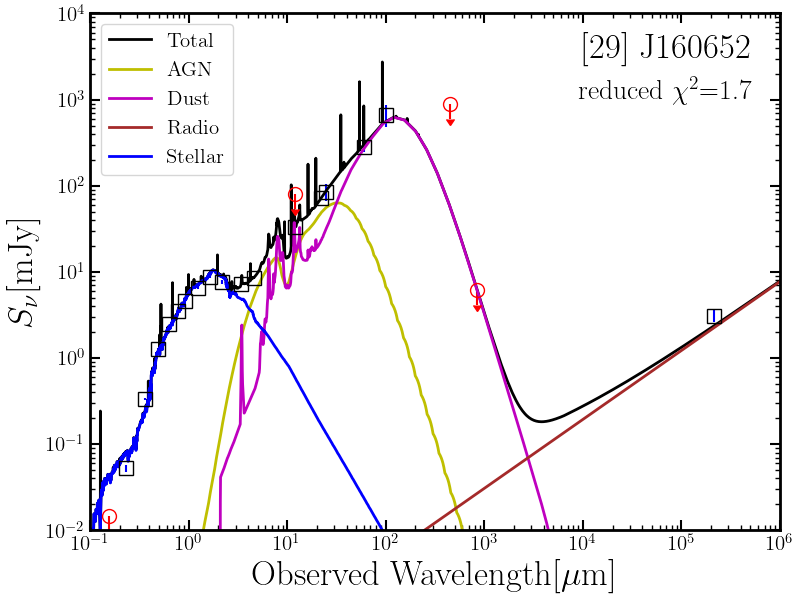}
\includegraphics[height=4.5cm, width=0.33\textwidth]{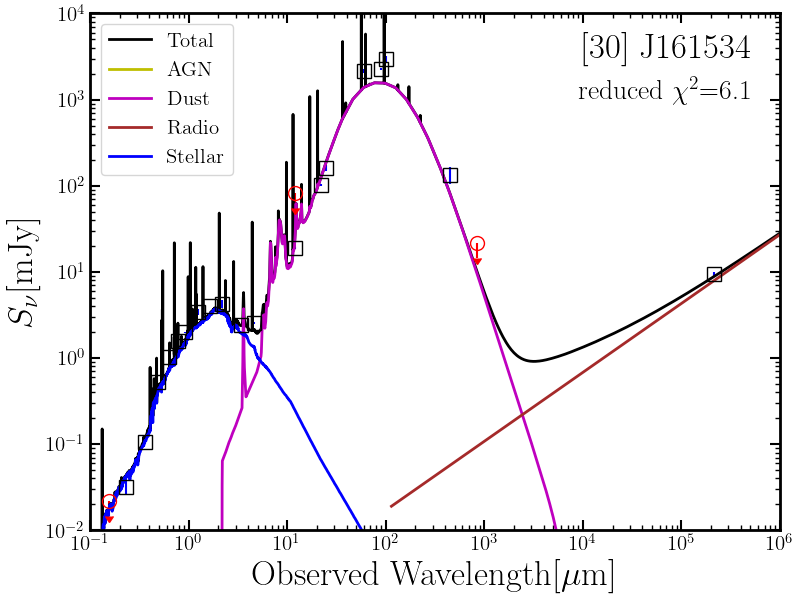}\\

\caption{\it continued }
\label{cigaleeg2}
\end{figure*}

\addtocounter{figure}{-1}
\begin{figure*}[htb!]
\includegraphics[height=4.5cm, width=0.33\textwidth]{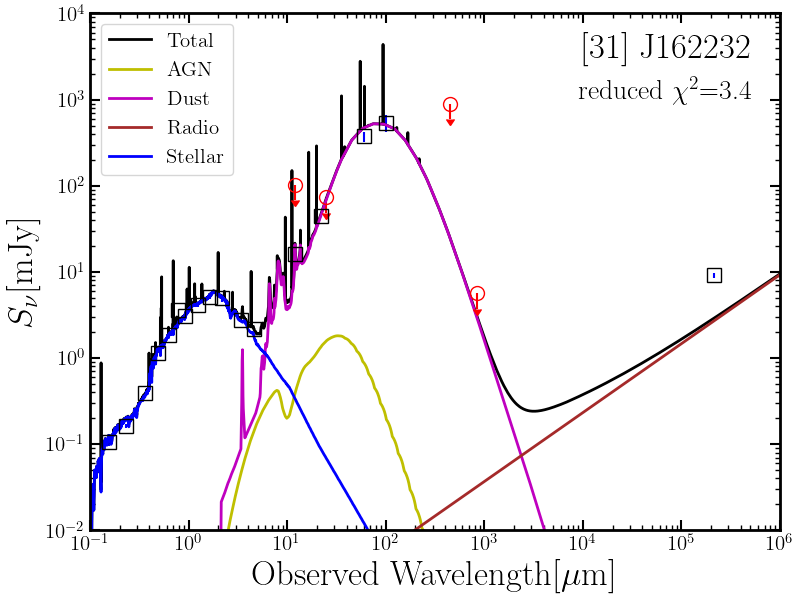}
\includegraphics[height=4.5cm, width=0.33\textwidth]{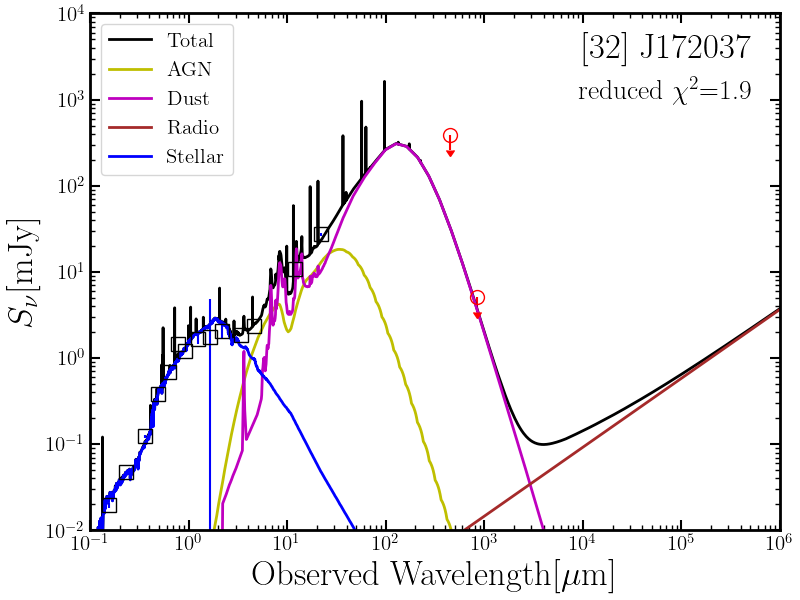}
\includegraphics[height=4.5cm, width=0.33\textwidth]{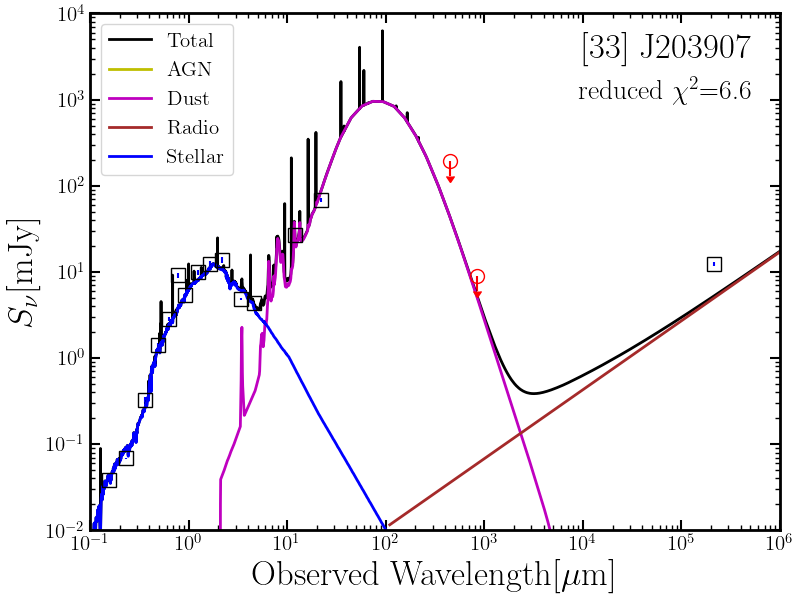}\\
\includegraphics[height=4.5cm, width=0.33\textwidth]{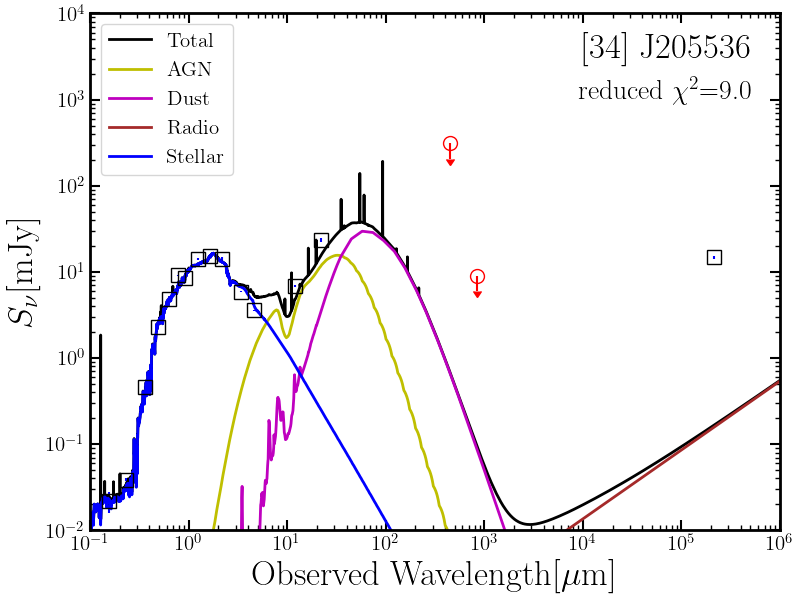}
\includegraphics[height=4.5cm, width=0.33\textwidth]{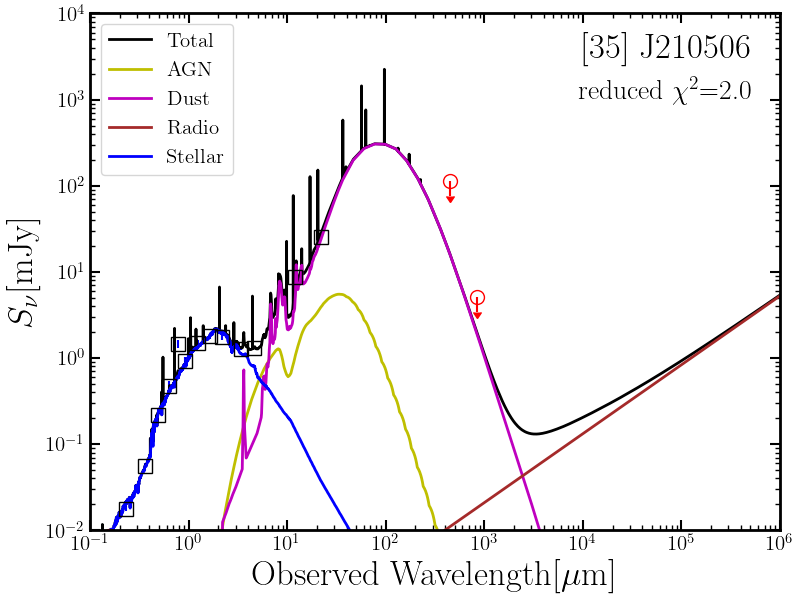}
\includegraphics[height=4.5cm, width=0.33\textwidth]{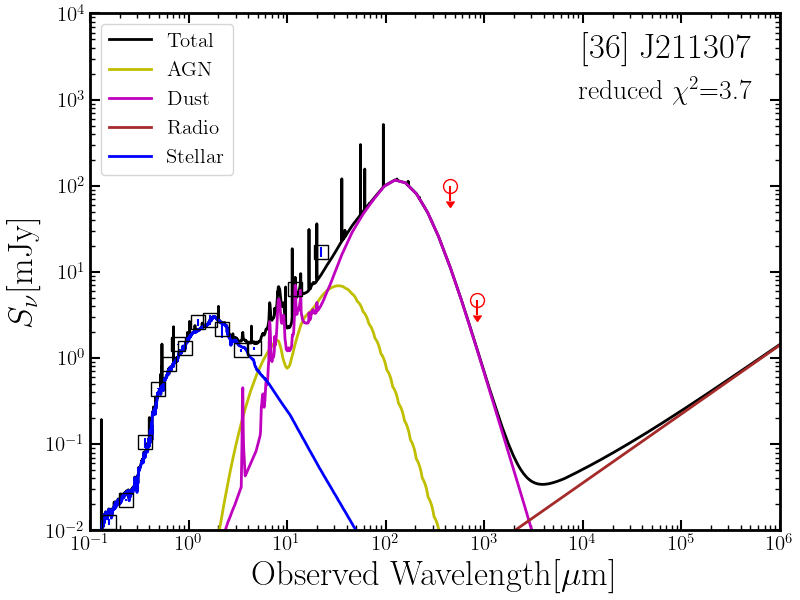}\\
\includegraphics[height=4.5cm, width=0.33\textwidth]{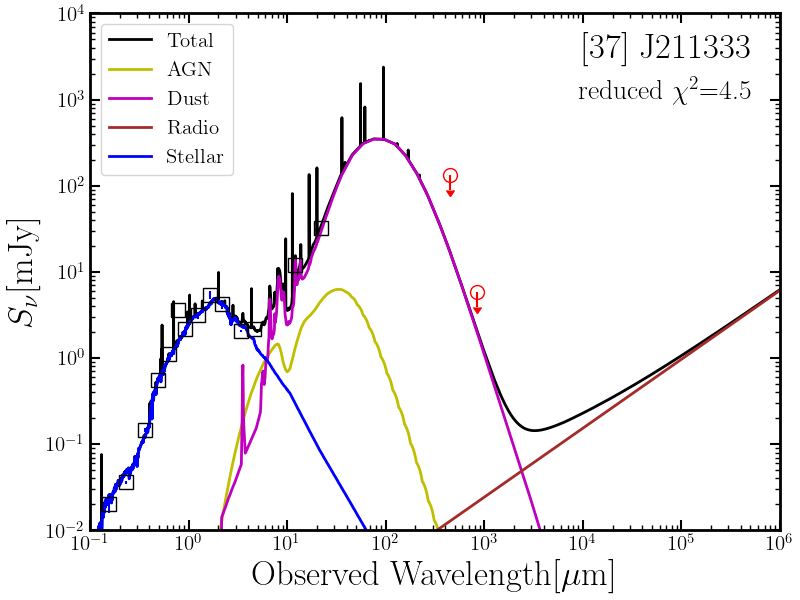}
\includegraphics[height=4.5cm, width=0.33\textwidth]{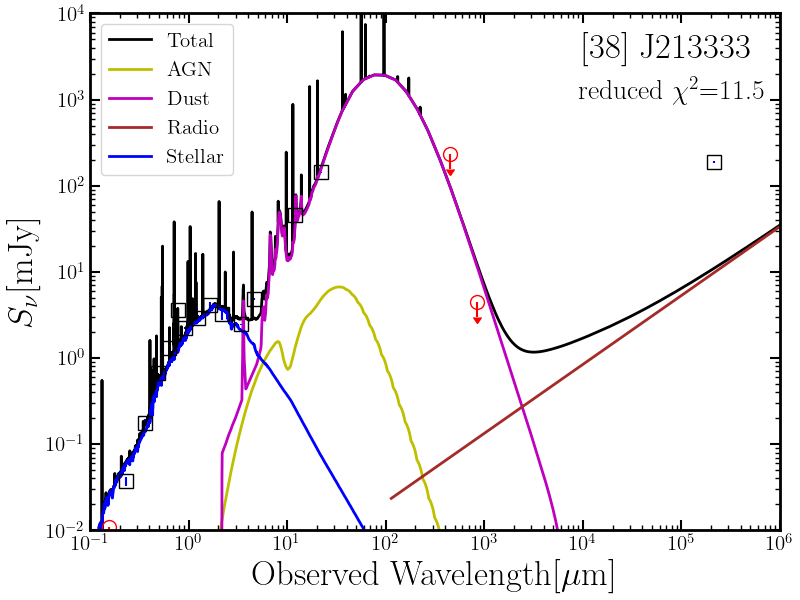}
\includegraphics[height=4.5cm, width=0.33\textwidth]{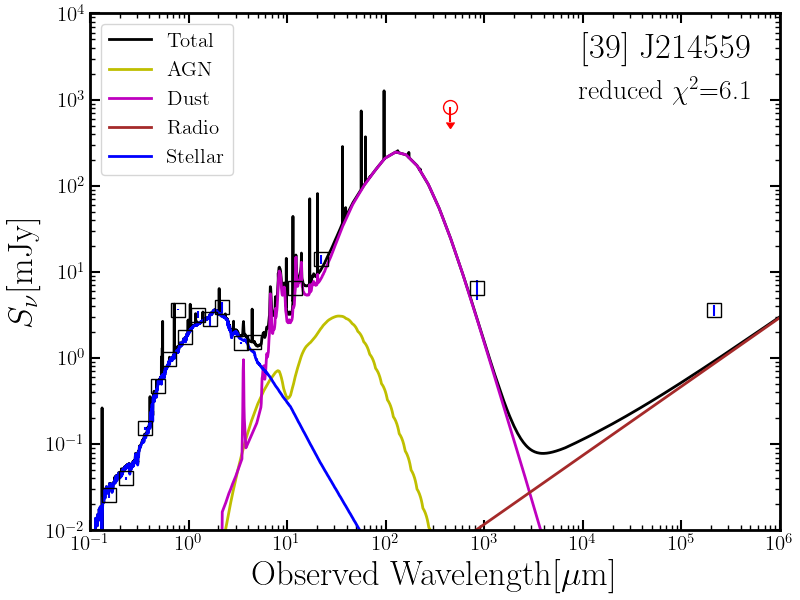}\\

\caption{ \it continued}
\label{cigaleeg3}
\end{figure*}
 
\begin{figure*}[htb!]
\includegraphics[height=4.5cm, width=0.33\textwidth]{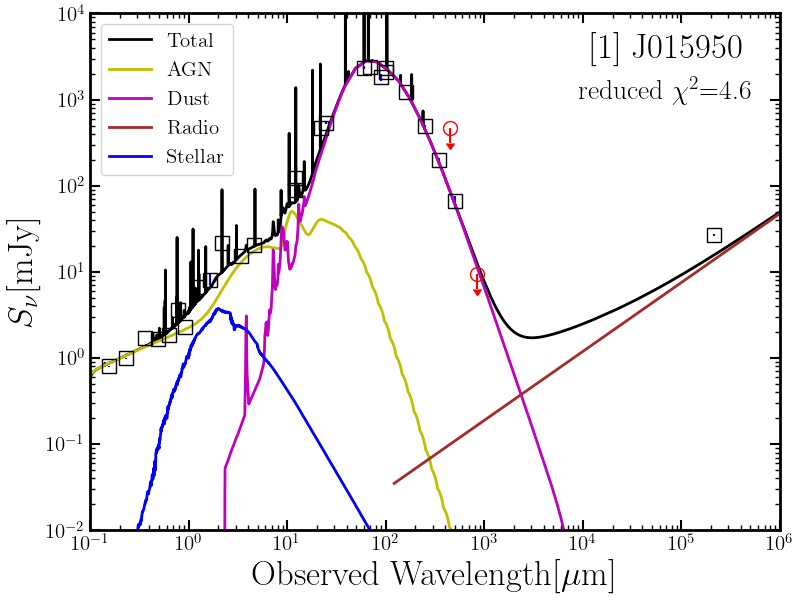}
\includegraphics[height=4.5cm, width=0.33\textwidth]{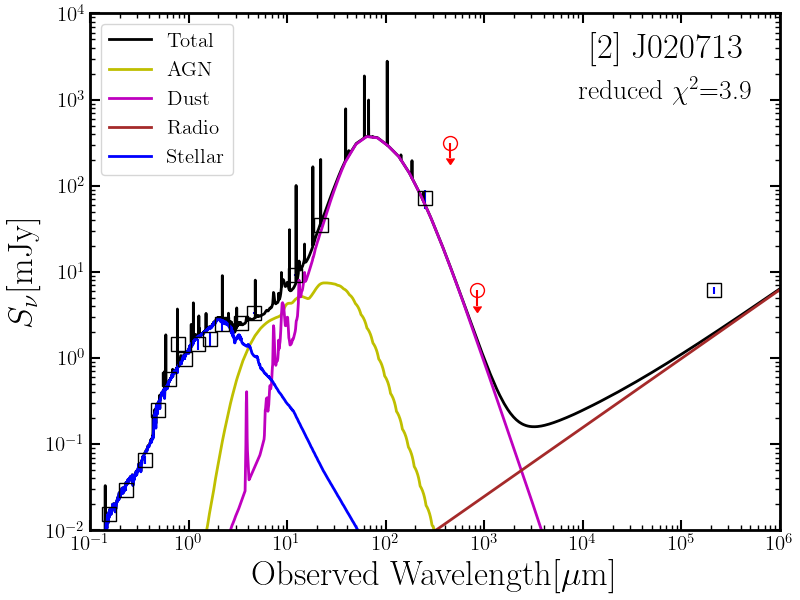}
\includegraphics[height=4.5cm, width=0.33\textwidth]{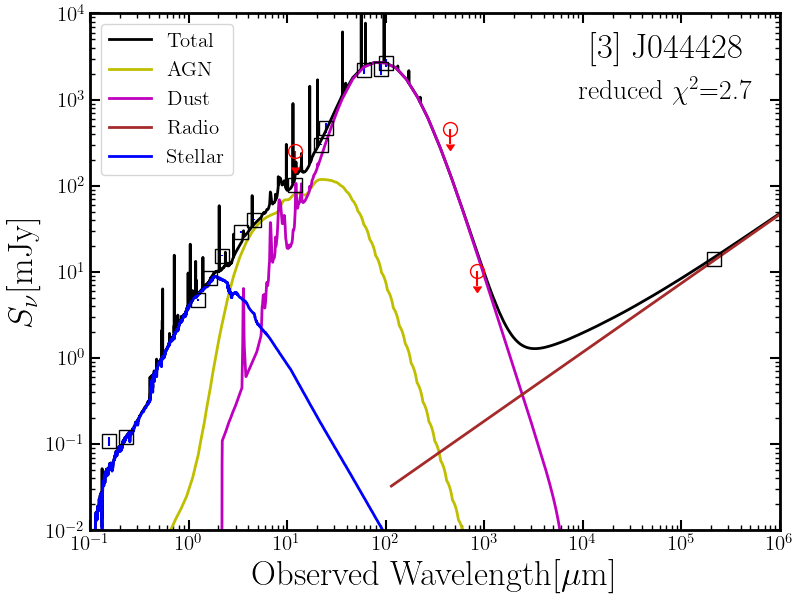}\\
\includegraphics[height=4.5cm, width=0.33\textwidth]{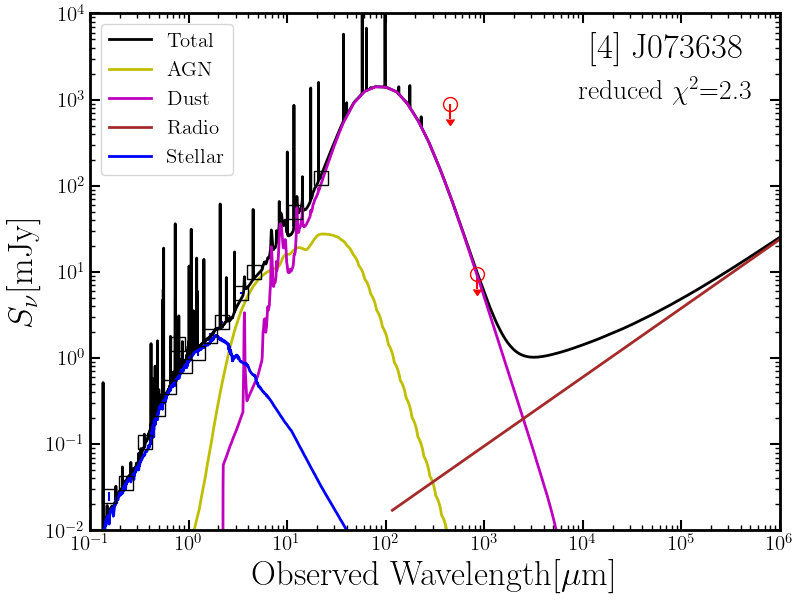}
\includegraphics[height=4.5cm, width=0.33\textwidth]{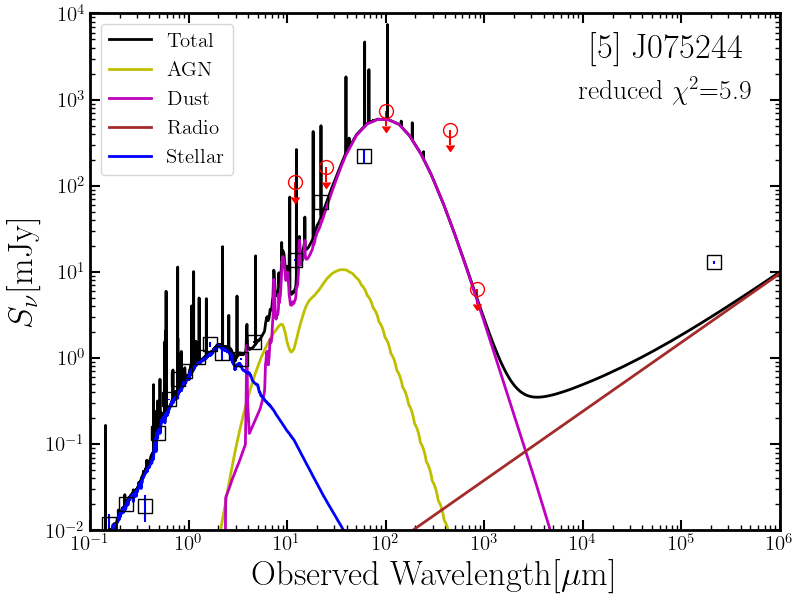}
\includegraphics[height=4.5cm, width=0.33\textwidth]{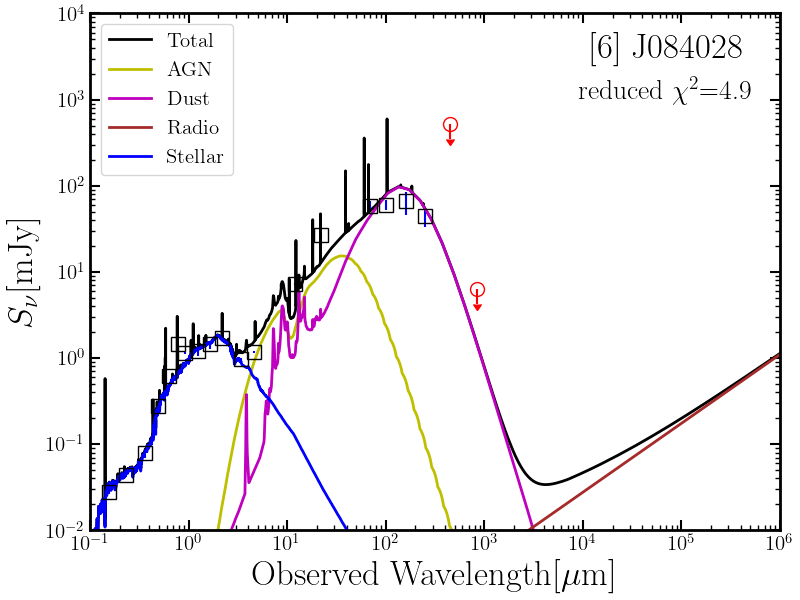}\\
\includegraphics[height=4.5cm, width=0.33\textwidth]{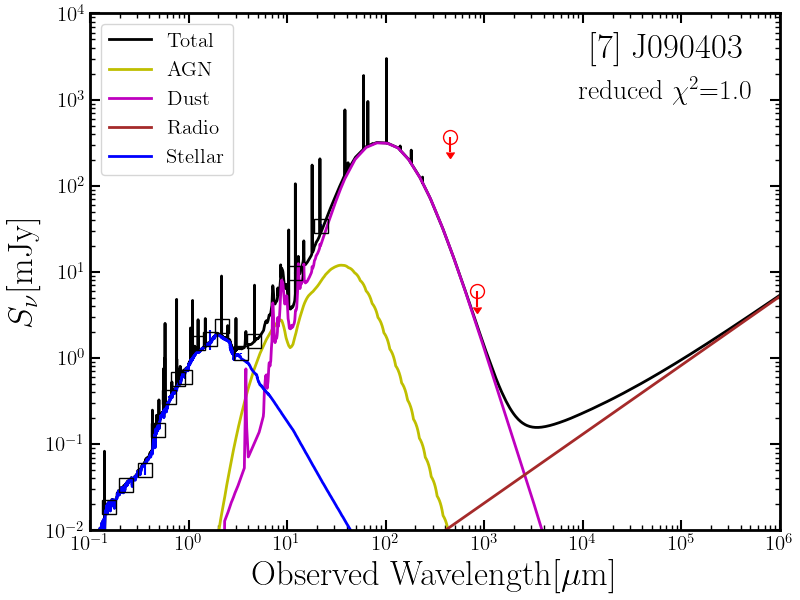}
\includegraphics[height=4.5cm, width=0.33\textwidth]{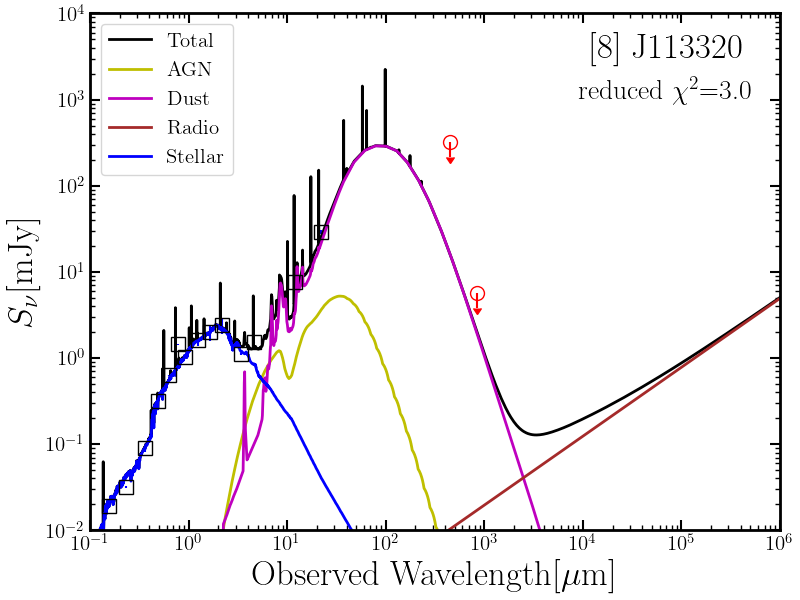}
\includegraphics[height=4.5cm, width=0.33\textwidth]{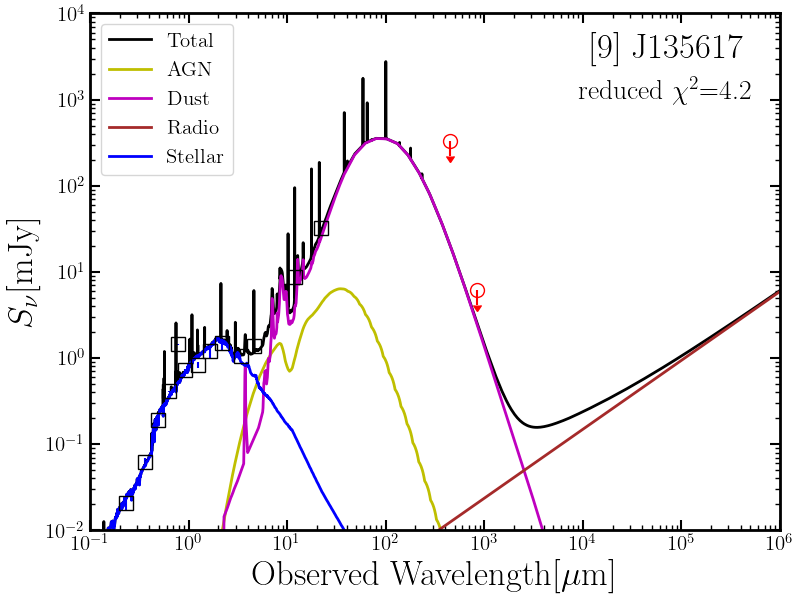}\\
\includegraphics[height=4.5cm, width=0.33\textwidth]{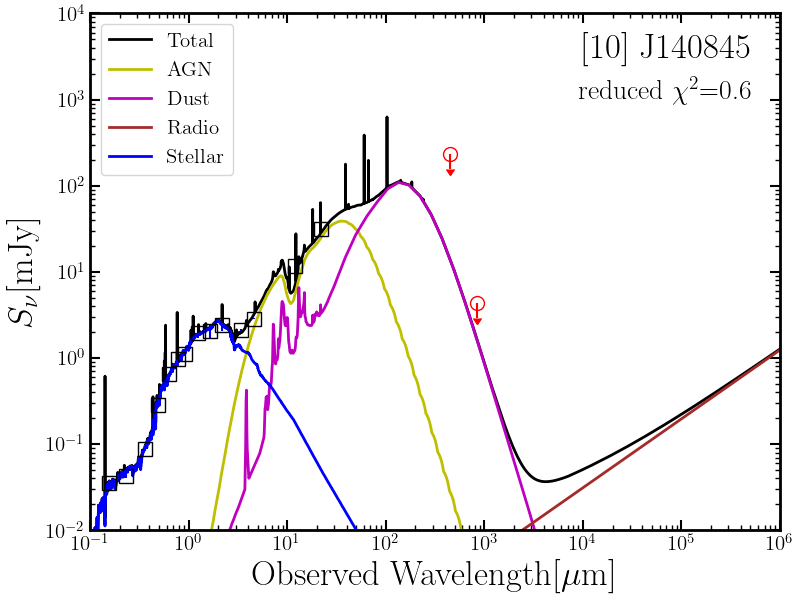}
\includegraphics[height=4.5cm, width=0.33\textwidth]{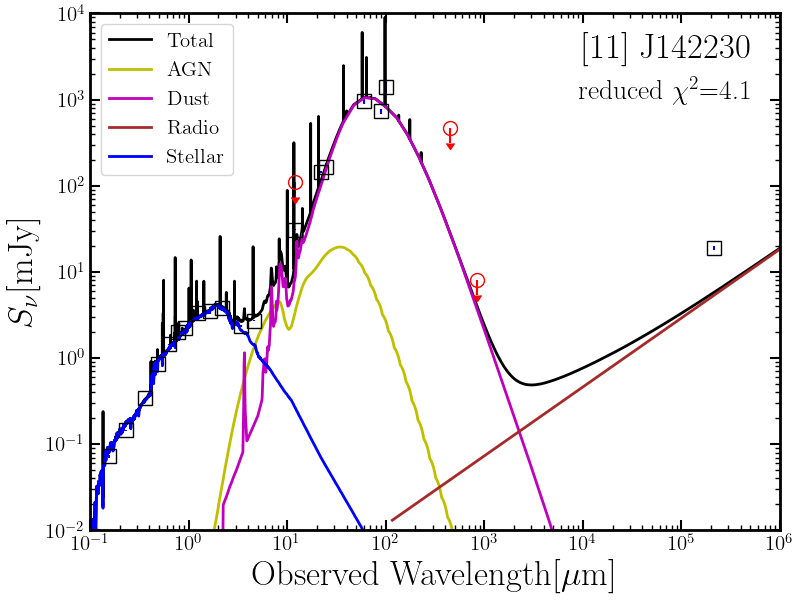}
\includegraphics[height=4.5cm, width=0.33\textwidth]{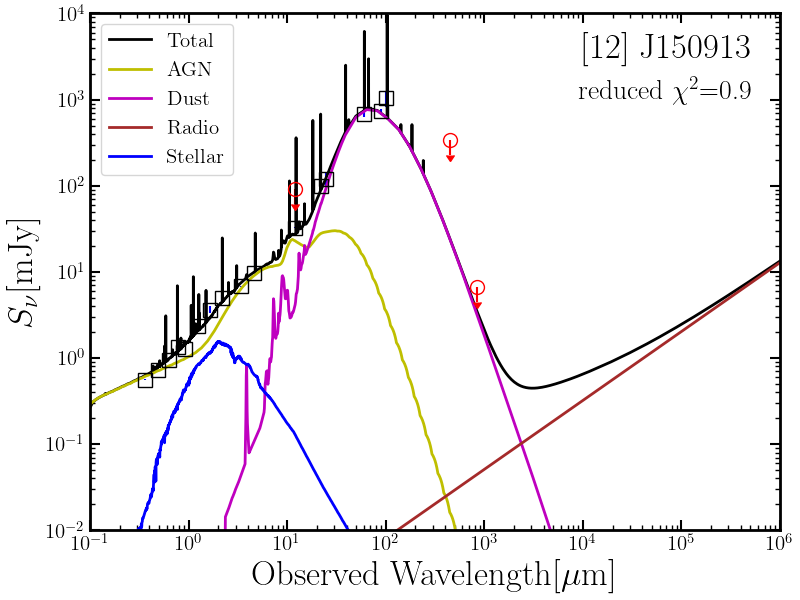}\\
\includegraphics[height=4.5cm, width=0.33\textwidth]{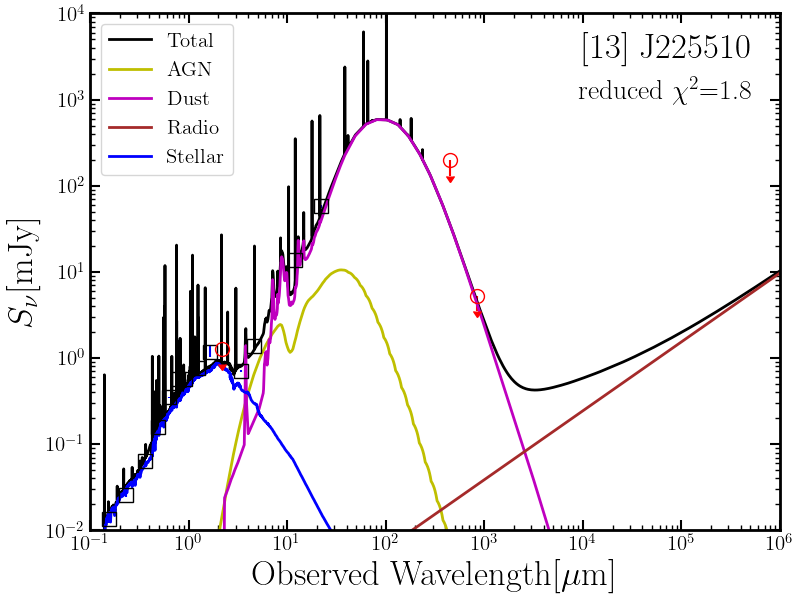}
\caption{The best-fit SED model (black line) of 13 type 1 AGNs. A dust component (magenta line) and an AGN component (green line) are also shown.}
\label{cigaleegt1}
\end{figure*}

\subsection{Multi-wavelength data} \label{subsec:mlambda}
To construct multi-wavelength data set, we collected flux data for each target from \citeauthor{NED}\footnote{https://ned.ipac.caltech.edu/}. UV (far-UV and near-UV) from Galaxy Evolution Explorer (GALEX), optical ($ugriz$) from SDSS, Near-IR ($J, H, K$ band) from Two Micron All Sky Survey (2MASS) Extended Source Catalog (XSC), Wide-field Infrared Survey Explorer (WISE) data from ALLWISE Source Catalog, Infrared Astronomical Satellite (IRAS) Faint Source Catalog, and 1.4 GHz from NRAO Very Large Array (VLA) Sky Survey (NVSS) \citep{Condon+98_NVSS} are included. In the case of FIR, we also utilized the database from the AKARI/Far-infrared Surveyor (FIS) all sky bright source catalog \citep{Yamamura:2010}, Herschel/PACS and the SPIRE point source catalog \citep{Marton:2016}.

\subsection{SED Analysis} \label{subsec:SED}

\newcolumntype{C}[1]{>{\centering\arraybackslash}m{#1}}
\begin{table*}[]
\centering
\caption{List of the input parameter space used in CIGALE}
\label{table_CIGALEpar}
\begin{tabular*}{\textwidth}{C{0.5\textwidth}C{0.5\textwidth}}
\toprule
\hline
Parameters & Input values \\
\hline
\multicolumn{2}{c}{Delayed SFH} \\
e-folding time of old stellar population (Myr) & 1, 500, 1000, 2000, 3000, 4000, 5000, 6000, 8000\\
e-folding time of young stellar population (Myr) & 20000\\
Age of old stellar population (Myr) & 10000\\
Age of young stellar population (Myr) & 5, 10, 50, 100, 300, 500, 1000\\
Mass fraction of young stellar population & 0.0, 0.01, 0.05, 0.1, 0.3, 0.5 \\
\hline
\multicolumn{2}{c}{Dust attenuation (Calzetti 2000 \& Leitherer 2002)} \\
$E(B-V)_*$ of stellar continuum by young population & 0.05, 0.1, 0.15, 0.2, 0.25, 0.3, 0.35, \\ &  0.4, 0.5, 0.6, 0.7, 0.8, 0.9 \\
$E(B-V)_*$ reduction factor of old stellar population & 0.44 \\ 
\hline
\multicolumn{2}{c}{Dust emission (Dale 2014)} \\
Slope alpha & 1.0, 1.5, 2.0, 2.5, 3.0 \\
\hline
\multicolumn{2}{c}{AGN emission (Skirtor 2016)} \\
Optical depth at 9.7 $\mu$m & 3, 7, 11 \\
Radial gradient of dust density & 1 \\
Angular gradient of dust density & 1 \\
Half opening angle of dusty torus & 40$^{\circ}$ \\
Ratio between inner/outer radii of torus & 20 \\
Inclination (i.e., viewing angle) & 30$^{\circ}$, 70$^{\circ}$ (type 1) \\ &  70$^{\circ}$ (type 2) \\
AGN fraction of $\mathrm{L_{IR}}$ & 0, 0.01, 0.05, 0.1,  0.2, 0.3, 0.4, 0.5, 0.6, 0.7 \\
\hline
\bottomrule 
\end{tabular*}
\end{table*}

We performed SED analysis using two different fitting methods. First, we used a python Code Investigating GAlaxy Emission {\citep[CIGALE\footnote{http://www.oamp.fr/cigale}][version 2020.0]{Noll:2009A&A,Boquien19}} to measure the IR luminosity emitted by dust in the ISM. CIGALE is a code for SED modeling of galaxies based on the energy balance principle, such that the emitted energy by dust in the IR range originates from the absorbed energy in the UV to NIR range. Based on the combinations of input parameters, CIGALE calculates all model grids and estimates physical parameters by Bayesian methods (exp(${-\chi^2/2}$)-weighted mean and error). It also provides a best-fit among all input models based on the minimum $\chi^2$. CIGALE covers a large wavelength range from UV to radio and provides multiple modules of each component in the SED. We briefly describe the adopted models in this paper as follows.

We adopt \citet{BC03} models for stellar populations, with a Chabrier initial mass function (IMF) for consistency with our previous works on SFR estimates \citep{Woo:2017ApJ, Woo:2020}. Metallicity is fixed as the solar metallicity (0.02). In nebular emission model, parameters are set as default values, ionization parameter (log U) as -2.0, the fraction of Lyman continuum photons escaping the galaxy and absorbed by dust ($f_{esc},\;f_{dust}$) as 0, and the line width as 300 \kms. For the star formation history (SFH), we adopt a delayed SFH model with two populations since this model reproduces better fits than normal single or double exponential models \citep{Ciesla:2015A&A}. Dust attenuation model is adopted from \citet{Calzetti:2000ApJ} and \citet{Leitherer02} (i.e., dustatt\_calzleit module) with $E(B-V)_*$ of young stellar population from 0.05 to 0.9. We fix the dust reddening reduction factor of older stellar population as 0.44. Since we focus on measuring dust luminosity, without constraining physical properties of dust, we adopted a simple dust emission model from \citet{Dale2014} which has only one parameter (i.e., slope $\alpha : dM_d(U) \propto U^{-\alpha}dU$). For the AGN component, we adopt SKIRTOR model \citep{Stalevski+12, Stalevski+16}, which has been recently updated in CIGALE and X-CIGALE \citep{Yang+20}. SKIRTOR model is more recommended than Fritz model \citep{Fritz:2006MNRAS} since it is based on a clumpy torus model and considers anisotropic emission from central source. Synchrotron radio emission is modelled by fixing radio-FIR correlation coefficient (qir) as 2.58 and the slope of the power law as 0.8. In summary, the total number of investigated models is $\approx$ 1.5 million (See Table \ref{table_CIGALEpar} for the full parameter space of CIGALE fitting).

Second, for a consistency check, we performed SED fits by using Chary-Elbaz templates \citep{Chary01}, which include a set of 105 IR SEDs constructed based on four typical galaxies in different luminosities. For each target, we chose the template and scale factor that give minimum $\chi^2$ for given FIR fluxes and upper limits. The upper limit of flux at the IR region is very useful to constrain the model and avoid overestimation of the scale factor of the SED. IR luminosities are calculated from the best-fit model by using the relation from Table 1 in \citet{Sanders+96}. Since these templates do not include hot dusty torus emission heated by AGN, the measured SFR is likely to be overestimated if AGN dust component is significant.  Thus, we adopted the CIGALE as a primary fitting method, and the comparison of the two methods are presented in Section 4.1. Note that we only used $\lambda > 20\;\mu$m IR data in Chary-Elbaz SED fitting to reduce the effect of AGN dust component in MIR.

We present the best-fit SED model along with the multi-wavelength data for 39 type 2 AGNs and 13 type 1 AGNs, respectively in Figure \ref{cigaleeg1} and Figure \ref{cigaleegt1}.  Note that the best-fit SED model is based on the reduced $\chi^2$ values from CIGALE for visual inspection and consistency check, while for SFR estimates, we adopted the IR luminosity determined based on the Bayesian analysis. In general we obtained good fits in the IR range. In particular, sub-mm data play an important role in the SED fit of the dust component when FIR data are unavailable. We found an unacceptable fit for one target, [18]J121156 (type 2), which is one of the interacting galaxies (see Figure \ref{Interactingeg}). Most IR excess is likely to originate from the interacting galaxy pair. Thus, the SFR based on the SED fit of this target is unreliable and excluded from the further analysis. 
Finally, we determine SFR from the IR luminosity (L$_\mathrm{IR}$) produced by dust using the calibration by \citet{Kennicutt:1998ARA&A} as revised for the Chabrier IMF:
\begin{equation}
\label{eq_SFR_LIR}
\centering
\rm{log\;SFR}\;(M_{\sun}\;yr^{-1}) = \rm{log}\; L_{IR}\;(erg\;s^{-1}) - 43.591.
\end{equation}

\section{Results and Discussion} \label{sec:results}

\begin{table*}[h]
\centering
\caption{Star formation rates and kinematic information for the type 2 AGNs}
\label{tablet2_SFR}

\scriptsize
\setlength{\tabcolsep}{4.5pt}
\begin{tabular*}{\textwidth}{@{}ccccccccccccc@{}}
\toprule
\hline

 & SDSS ID  & SFR$_{\rm{IR}}$         & SFR$_{\rm{Dn4000}}$      & SFR$_{\rm{ANN}}$       & M$_{\star}$       & $\sigma_{\star}$  & $\sigma_{\rm{[OIII]}}$ & $V_{\rm{[OIII]}}$ & L$_{\rm{[OIII]}}$ & M$_{BH}$           & E.R.       & frac$_{\rm{AGN}}$ \\

 &          & [M$_{\sun}/year$]  & [M$_{\sun}/year$]   & [M$_{\sun}/year$] & [M$_{\sun}$]      & [km/s]            &  [km/s]           &       [km/s]       &[erg/s]       & [M$_{\sun}$]       &            &\\
                       
        {[}1{]}            & {[}2{]}  & {[}3{]}  & {[}4{]}  & {[}5{]}     & {[}6{]}           & {[}7{]}      & {[}8{]}          & {[}9{]}       & {[}10{]}        & {[}11{]}    &   {[}12{]}  &   {[}13{]}  \\ \midrule 

1	   &  J075328.31+142141.0 	&  0.88$\pm$0.02    &  0.99$\pm$0.30    & 1.38$\pm$0.10  & 10.85     & 132$\pm$8     & 390$\pm$9	     &	 1$\pm$12    & 42.31	 & 8.06	    & -1.62	    & 0.00$\pm$0.00 \\
2      &  J082714.96+263618.2 	&  0.95$\pm$0.16    &  0.90$\pm$0.42    & 1.20$\pm$0.03  & 10.76     & 148$\pm$10    & 379$\pm$15        & -121$\pm$19   & 40.94     & 7.97     & -1.59     & 0.30$\pm$0.14\\
3      &  J083132.28+160143.3 	&  0.09$\pm$0.10    &  0.38$\pm$0.28    & 0.52$\pm$0.04  & 10.22     &  91$\pm$9     & 385$\pm$28        & -202$\pm$21   & 40.64     & 7.40     & -1.31     & 0.38$\pm$0.06   \\
4      &  J084344.98+354942.0 	&  0.62$\pm$0.02    &  0.95$\pm$0.45    & -              & 11.07     & 192$\pm$8     & 190$\pm$1         &  -46$\pm$6    & 41.74     & 8.30     & -1.11     & 0.30$\pm$0.01   \\
5      &  J085547.66+004739.4 	&  0.90$\pm$0.02    &  0.63$\pm$0.33    & 0.90$\pm$0.09  & 10.96     & 130$\pm$5     & 255$\pm$25        & -230$\pm$15   & 40.41     & 8.18     & -2.32     & 0.00$\pm$0.01   \\
6      &  J091807.52+343946.0 	&  1.33$\pm$0.02    &  1.60$\pm$0.43    & 1.69$\pm$0.06  & 11.01     & 138$\pm$10    & 259$\pm$10        & -261$\pm$17   & 40.86     & 8.23     & -1.93     & 0.00$\pm$0.01   \\
7      &  J101936.79+193313.4 	&  0.14$\pm$0.12    & -0.37$\pm$0.50    & 0.39$\pm$0.07  & 10.51     & 136$\pm$6     & 132$\pm$4         &  -17$\pm$11   & 41.39     & 7.71     & -0.87     & 0.37$\pm$0.10   \\
8      &  J103723.62+021845.5 	&  1.11$\pm$0.02    & -1.12$\pm$0.70    & -0.31$\pm$0.06 & 10.75     & 198$\pm$5     & 190$\pm$17        &    4$\pm$23   & 40.05     & 7.96     & -2.46     & 0.01$\pm$0.02   \\
9      &  J104232.05+050241.9 	&  0.44$\pm$0.02    & -1.19$\pm$0.51    & -              & 10.60     &  84$\pm$7     &  82$\pm$6         &   16$\pm$10   & 39.98     & 6.30     & -0.87     & 0.10$\pm$0.00   \\
10     &  J105833.33+461604.8 	&  0.02$\pm$0.02    & -0.32$\pm$0.70    & -              & 11.06     & 142$\pm$5     & 139$\pm$4         &   31$\pm$10   & 40.89     & 8.29     & -1.95     & 0.00$\pm$0.00   \\
11     &  J110037.22+112455.1 	&  0.04$\pm$0.02    & -0.06$\pm$0.58    & -              & 10.97     & 165$\pm$5     & 199$\pm$13        &  115$\pm$24   & 39.55     & 8.20     & -3.20     & 0.01$\pm$0.01   \\
12	   &  J110630.64+063333.9 	&  0.73$\pm$0.02    &  0.69$\pm$0.31	& 0.64$\pm$0.04  & 10.70     & 134$\pm$6	 & 174$\pm$2         & -157$\pm$7    & 42.01	 & 7.90     & -1.69     & 0.06$\pm$0.02   \\
13     &  J111406.30+554239.1 	&  0.79$\pm$0.02    &  0.58$\pm$0.38    & 0.73$\pm$0.05  & 10.71     & 131$\pm$5     & 155$\pm$6         &  -34$\pm$12   & 39.76     & 7.92     & -2.71     & 0.00$\pm$0.00   \\
14     &  J113549.07+565708.2 	&  1.36$\pm$0.03    & 1.16$\pm$0.32     & -              & 11.03     & 207$\pm$7     & 400$\pm$9         & -213$\pm$11   & 41.69     & 8.26     & -1.12     & 0.35$\pm$0.05   \\
15     &  J113606.63+621456.9 	&  0.99$\pm$0.02    & 0.51$\pm$0.33     & -              & 10.90     & 157$\pm$5     & 168$\pm$4         &. -31$\pm$8    & 40.75     & 8.12     & -1.93     & 0.00$\pm$0.00    \\
16	   &  J114719.93+075243.0 	&  1.23$\pm$0.06	& 0.67$\pm$0.40     & -              & -         & 133$\pm$9     & 476$\pm$18	     & -344$\pm$16	 & 42.41     & -		& -         & 0.08$\pm$0.03   \\
17     &  J115657.88+550821.5 	&  0.56$\pm$0.05    & -0.02$\pm$0.58    & 0.61$\pm$0.07  & 11.02     & 160$\pm$7     & 156$\pm$2         &   32$\pm$6    & 41.20     & 8.25     & -1.60     & 0.00$\pm$0.01    \\
18     &  J121155.63+372113.4 	&  0.45$\pm$0.16    & 1.41$\pm$0.69     & 0.50$\pm$0.10  & 11.17     & 131$\pm$9     & 223$\pm$9         &  -19$\pm$16   & 41.12     & 8.05     & -1.48     & 0.06$\pm$0.06   \\
19     &  J125642.71+350729.9 	&  1.44$\pm$0.02    & 0.69$\pm$0.42     & 0.74$\pm$0.08  & 10.44     & 177$\pm$13    & 288$\pm$163       & -49$\pm$76    & 39.71     & 7.63     & -2.47     & 0.00$\pm$0.00     \\
20     &  J131153.80+053138.3 	&  0.44$\pm$0.18    & 0.45$\pm$0.36     & 0.82$\pm$0.10  & 10.93     & 175$\pm$7     & 172$\pm$3         & 39$\pm$12     & 41.48     & 8.16     & -1.23     & 0.60$\pm$0.09    \\
21     &  J132948.19+310748.5 	&  0.53$\pm$0.02    & 0.50$\pm$0.31     & -              & 10.76     & 107$\pm$4     & 148$\pm$23        & -13$\pm$24    & 39.45     & 7.97     & -3.07     & 0.00$\pm$0.00   \\
22     &  J140452.65+532332.1 	&  1.09$\pm$0.08    & 0.95$\pm$0.32     & -              & 10.80     & 191$\pm$9     & 392$\pm$8         & -303$\pm$12   & 41.44     & 8.01     & -1.13     & 0.15$\pm$0.05    \\  
23	   &  J141631.74+393521.2 	&  0.85$\pm$0.02    & 0.37$\pm$0.32     &-	             & 10.77	 & 129$\pm$4     &163$\pm$120        & -24$\pm$76    & 39.93	 & 7.98	    & -3.20     & 0.00$\pm$0.00   \\
24     &  J142859.53+605000.5 	&  1.26$\pm$0.02    & 0.87$\pm$0.67     & 1.36$\pm$0.10  & 10.97     & 148$\pm$5     & 247$\pm$76        & -153$\pm$44   & 40.23     & 8.19     & -2.52     & 0.00$\pm$0.00   \\  
25     &  J143545.74+244332.8 	&  0.98$\pm$0.02    & 0.99$\pm$0.39     & -              & 11.21     & 169$\pm$6     & 180$\pm$14        & -114$\pm$20   & 40.62     & 8.44     & -2.38     & 0.00$\pm$0.00   \\  
26     &  J145835.98+445300.9 	&  1.32$\pm$0.02    & 1.07$\pm$0.34     & -              & 11.16     & 177$\pm$5     & 312$\pm$14        & 101$\pm$15    & 40.83     & 8.40     & -2.12     & 0.01$\pm$0.01   \\  
27     &  J152549.54+052248.7 	&  1.57$\pm$0.05    & 1.22$\pm$0.37     & 1.37$\pm$0.05  & 10.79     & 183$\pm$9     & 257$\pm$10        & -69$\pm$17    & 40.27     & 8.00     & -2.28     & 0.00$\pm$0.00   \\
28     &  J160507.88+174527.6 	&  0.79$\pm$0.03    & -0.53$\pm$0.68    & -              & 11.28     & 177$\pm$4     & 121$\pm$18        & 30$\pm$23     & 39.21     & 8.53     & -3.87     & 0.00$\pm$0.00   \\
29     &  J160652.16+275539.1 	&  0.62$\pm$0.11    & 0.56$\pm$0.36     & 0.76$\pm$0.06  & 10.61     & 128$\pm$6     & 324$\pm$7         & -268$\pm$15   & 40.85     & 7.81     & -1.51     & 0.29$\pm$0.08   \\
30     &  J161534.13+210019.7 	&  1.68$\pm$0.06    & 1.10$\pm$0.47     & 1.36$\pm$0.06  & 10.79     & 164$\pm$12    & 210$\pm$18        & 16$\pm$21     & 40.30     & 8.00     & -2.26     & 0.00$\pm$0.00   \\
31     &  J162232.68+395650.2 	&  0.96$\pm$0.02    & 0.94$\pm$0.29     & 1.34$\pm$0.06  & 10.60     & 142$\pm$8     & 556$\pm$11        & -34$\pm$17    & 41.57     & 7.80     & -0.78     & 0.01$\pm$0.01    \\
32     &  J172037.94+294112.4 	&  1.00$\pm$0.07    & 0.77$\pm$0.28     & 1.09$\pm$0.04  & 10.61     & 116$\pm$9     & 401$\pm$23        & -65$\pm$14    & 41.30     & 7.82     & -1.07     & 0.19$\pm$0.05   \\
33	   &  J203907.05+003316.3 	&  0.98$\pm$0.02    & 0.62$\pm$0.36     & 0.80$\pm$0.03  & 10.79     & 151$\pm$5	 & 508$\pm$32        & -47$\pm$22    & 42.14     & 8.00	    & -1.63	    & 0.00$\pm$0.00   \\
34     &  J205536.51-003811.7 	& -0.42$\pm$0.06    & 0.60$\pm$0.70     & -              & 11.09     & 191$\pm$6     & 245$\pm$10        & 206$\pm$10    & 40.93     & 8.32     & -1.94     & 0.59$\pm$0.03   \\
35     &  J210506.94+094118.8 	&  1.05$\pm$0.07    & 1.21$\pm$0.35     & 1.47$\pm$0.09  & 10.70     & 144$\pm$11    & 384$\pm$27        & -145$\pm$29   & 40.99     & 7.90     & -1.47     & 0.06$\pm$0.03   \\
36     &  J211307.21+005108.4 	&  0.39$\pm$0.19    & 0.41$\pm$0.33     & 0.59$\pm$0.11  & 10.40     & 84$\pm$10     & 374$\pm$22        & -95$\pm$15    & 40.81     & 7.59     & -1.33     & 0.12$\pm$0.08   \\
37     &  J211333.79-000950.4 	&  0.93$\pm$0.07    & 0.57$\pm$0.32     & 1.02$\pm$0.04  & 10.72     & 136$\pm$8     & 384$\pm$7         & -56$\pm$13    & 41.20     & 7.93     & -1.28     & 0.03$\pm$0.03   \\
38     &  J213333.31-071249.2 	&  1.79$\pm$0.03    & 0.92$\pm$0.32     & 0.94$\pm$0.10  & 10.96     & 153$\pm$7     & 380$\pm$6         & -135$\pm$11   & 41.43     & 8.18     & -1.30     & 0.01$\pm$0.02   \\
39     &  J214559.99+111325.9 	&  0.78$\pm$0.07    & 0.60$\pm$0.29     & 1.12$\pm$0.04  & 10.79     & 142$\pm$9     & 394$\pm$8         & -39$\pm$12    & 41.54     & 8.00     & -1.01     & 0.03$\pm$0.04   \\      \bottomrule \hline

\end{tabular*}%
 \tablecomments{(1) Target number; (2) SDSS name; (3) Logarithm of SFR based on the IR luminosity, (4) Logarithm of SFR based on the D$_n$4000; (5) Logarithm of SFR based on artificial neural network method \citep{Ellison16a}; (6) Logarithm of stellar mass; (7) Stellar velocity dispersion; (8) Velocity dispersion of [\OIII] $\lambda$5007; (9) Velocity offset of [\OIII]$\lambda$5007; (10) Logarithm of [\OIII] luminosity; (11) Logarithm of black hole mass; (12) Logarithm of Eddington ratio. Note that kinematical properties, black hole mass, and Eddington ratio are adopted from \citet{Woo+16}.; (13) AGN fraction of IR luminosity.}
\end{table*}

\begin{table*}[]
\centering
\caption{Star formation rates and kinematic information for the type 1 AGNs}
\label{tablet1_SFR}
\scriptsize
\setlength{\tabcolsep}{4.5pt}
\begin{tabular*}{\textwidth}{@{}ccccccccccccc@{}}
\toprule
\hline

 & SDSS ID  & SFR$_{\rm{IR}}$         & SFR$_{\rm{Dn4000}}$      & SFR$_{\rm{ANN}}$       & M$_{\star}$       & $\sigma_{\star}$  & $\sigma_{\rm{[OIII]}}$ & $V_{\rm{[OIII]}}$ & L$_{\rm{[OIII]}}$ & M$_{BH}$           & E.R.       & frac$_{\rm{AGN}}$ \\

 &          & [M$_{\sun}/year$]  & [M$_{\sun}/year$]   & [M$_{\sun}/year$] & [M$_{\sun}$]      & [km/s]            &  [km/s]           &       [km/s]       &[erg/s]       & [M$_{\sun}$]       &            &\\
                       
        {[}1{]}  & {[}2{]}  & {[}3{]}  & {[}4{]}  & {[}5{]}     & {[}6{]}     & {[}7{]}      & {[}8{]}  & {[}9{]}       & {[}10{]}        & {[}11{]}    &   {[}12{]}  &   {[}13{]}  \\ \midrule 

1      & J015950.23+002340.9   	& 2.55$\pm$0.02  	& 1.66$\pm$0.30       & -		        & 11.47     & 179$\pm$47    & 527$\pm$10     & -243$\pm$12    & 42.79   & 8.24     & -0.68     & 0.10$\pm$0.00    \\
2      & J020713.32-011223.1   	& 1.70$\pm$0.05  	& 0.55$\pm$0.36       & 2.19$\pm$0.18   & 11.06     & 110$\pm$55    & 395$\pm$18     & -192$\pm$15    & 42.10   & 8.33     & -1.64     & 0.11$\pm$0.03    \\
3      & J044428.77+122111.7   	& 1.96$\pm$0.03  	& 1.35$\pm$0.21       & -		        & 11.07     & 151$\pm$104   & 490$\pm$55     & -559$\pm$15    & 42.19   & 8.11     & -0.74     & 0.20$\pm$0.02    \\
4      & J073638.86+435316.5   	& 1.82$\pm$0.12  	& 0.69$\pm$0.36       & 1.90$\pm$0.17   & 10.69     & 95$\pm$39     & 447$\pm$6      & -6$\pm$12      & 42.12   & 6.63     & -0.30     & 0.18$\pm$0.10    \\
5      & J075244.63+434105.3   	& 1.94$\pm$0.02  	& 0.16$\pm$0.59       & 2.17$\pm$0.27   & 11.15     & 155$\pm$90    & 411$\pm$24     & 63$\pm$15      & 42.15   & 6.55     & -0.25     & 0.05$\pm$0.01    \\
6      & J084028.61+332052.2   	& 0.94$\pm$0.02     & 0.29$\pm$0.39       & 1.71$\pm$0.15   & 11.19     & 199$\pm$78    & 445$\pm$38     & -31$\pm$13     & 42.19   & 6.30     & -0.51     & 0.39$\pm$0.03    \\
7      & J090403.72+074819.3   	& 1.56$\pm$0.07  	& 0.09$\pm$0.29       & 2.05$\pm$0.26   & 10.80     & 196$\pm$80    & 407$\pm$18     & -49$\pm$8      & 42.35   & 5.78     & -0.17     & 0.08$\pm$0.03    \\
8      & J113320.56-033337.5   	& 1.27$\pm$0.08  	& 0.02$\pm$0.60       & 1.26$\pm$0.18   & 10.87     & 94$\pm$43     & 414$\pm$8      & -23$\pm$12     & 42.01   & 6.52     & -0.34     & 0.05$\pm$0.04    \\
9      & J135617.79-023101.5   	& 1.45$\pm$0.03  	& 0.25$\pm$0.25       & 2.12$\pm$0.18   & 10.78     & 100$\pm$38    & 327$\pm$22     & 1$\pm$7        & 42.21   & 6.00     & -0.05     & 0.06$\pm$0.02    \\
10     & J140845.73+353218.4  	& 1.04$\pm$0.23		& 0.67$\pm$0.26       & 2.37$\pm$0.26   & 11.12     & 164$\pm$75    & 358$\pm$28     & 162$\pm$18     & 42.18   & 6.17     & 0.71      & 0.58$\pm$0.14    \\
11     & J142230.34+295224.2    & 1.79$\pm$0.02  	& 0.96$\pm$0.24       & -		        & 10.90     & 138$\pm$58    & 519$\pm$32     & -139$\pm$29    & 42.18   & 7.97     & -1.14     & 0.05$\pm$0.01    \\
12     & J150913.79+175710.0  	& 2.02$\pm$0.02 	& 2.73$\pm$0.02       & -		        & 11.10     & 179$\pm$147   & 367$\pm$17     & -107$\pm$22    & 42.42   & 7.95     & -0.72     & 0.20$\pm$0.00    \\
13     & J225510.11-081234.2   	& 1.84$\pm$0.05  	& 0.20$\pm$0.25       & 1.61$\pm$0.16   & 10.82     & 89$\pm$45     & 413$\pm$25     & 249$\pm$10     & 42.17   & 6.20     & 0.12      & 0.02$\pm$0.02    \\   \bottomrule 
\hline

\end{tabular*}%


 \tablecomments{(1) Target number; (2) SDSS name; (3) Logarithm of SFR based on the IR luminosity, (4) Logarithm of SFR based on the D$_n$4000; (5) Logarithm of SFR based on artificial neural network method \citep{Ellison16a}; (6) Logarithm of stellar mass; (7) Stellar velocity dispersion; (8) Velocity dispersion of [\OIII] $\lambda$5007; (9) Velocity offset of [\OIII]$\lambda$5007; (10) Logarithm of [\OIII] luminosity; (11) Logarithm of black hole mass; (12) Logarithm of Eddington ratio. Note that kinematical properties, black hole mass, and Eddington ratio are adopted from \citet{Rakshit18}.; (13) AGN fraction of IR luminosity.}

\end{table*}

\subsection{Dust luminosity based on the SED fit}\label{subsec:CE01comparison}

\begin{figure}[h]
	\includegraphics[width=1.05\columnwidth]{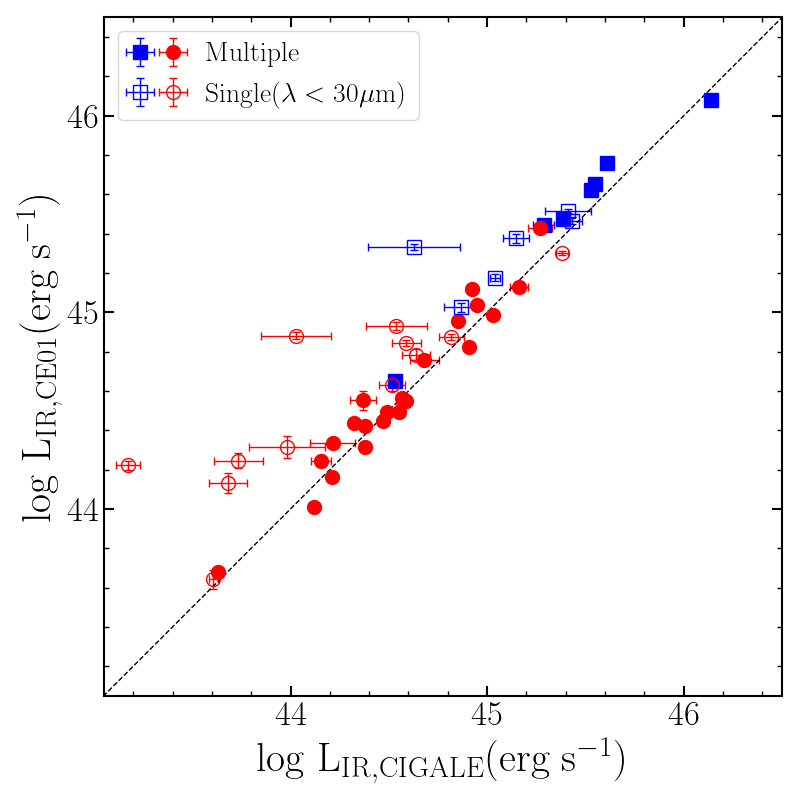}
	\caption{Comparison of the dust luminosities, respectively obtained by the CIGALE and Chary-Elbaz methods. Blue squares and red circles represent type 1 and type 2 respectively, while dashed line shows a one-to-one relation. Open symbols indicate targets with only one detection in MIR at $\lambda <30\;\mu$m.}
	\label{CE01comp}
\end{figure}

We measured the total IR luminosity re-emitted by dust based on the SED fit with the CIGALE method, after removing the contribution from AGN dust heating. The measured dust luminosity ranges over three orders of magnitude (i.e., 10$^{43}$ to 10$^{46}$ \ergs), covering various levels of star formation in the host galaxies. 

For a consistency check, we compare the dust luminosity measured using the CIGALE method with that obtained using the Chary-Elbaz method in Figure \ref{CE01comp}. The two methods provide consistent results within a factor of two as the scatter between the two measurements is 0.27 dex. However, there is a systematic difference with an average offset of 0.14 dex, indicating that the IR luminosity measured with the Chary-Elbaz method is higher than that obtained with the CIGALE by a factor of $\sim$1.4. This trend is consistent with the expectation that the Chary-Elbaz method overestimates dust luminosity due to the contribution from AGN dust heating. In contrast, AGN component is decomposed and removed during the SED fitting with the CIGALE method. 
Depending on the availability of FIR flux measurements, the systematic uncertainty of the Chary-Elbaz method can be larger. For example, when only one MIR flux point is detected, the overestimation is severer since the AGN dust heating is not well constrained as AGN dust mainly contributes to MIR. Consequently, if we only counted targets that were fit with one MIR data point at $\lambda < 30\;\mu$m, the systematic offset and scatter become larger as 0.29 and 0.41 dex, respectively. This offset is found even if frac$_{\rm{AGN}}$ is low since MIR AGN contamination can still be high in the case of low frac$_{\rm{AGN}}$ in total IR luminosity. Therefore, CE01 fitting of AGN with one MIR point can overestimate IR luminosity $\sim$ 0.29 dex.
In contrast, if we focus on the targets that have additional FIR or sub-mm flux measurements at $\lambda > 30\;\mu$m, the dust luminosities measured with CIGALE and Chary-Elbaz methods are consistent with a 0.05 dex offset and a 0.10 dex scatter. Thus, we conclude that the two methods provide consistent results on dust luminosity when FIR flux measurements are available, while there is a systematic uncertainty in using the Chary-Elbaz method, particularly when FIR data are insufficient. 

\subsection{Comparison with stellar population model based SFR}\label{comp_CIGALE_SFRs}
\begin{figure*}[t]

\includegraphics[width=0.5\textwidth]{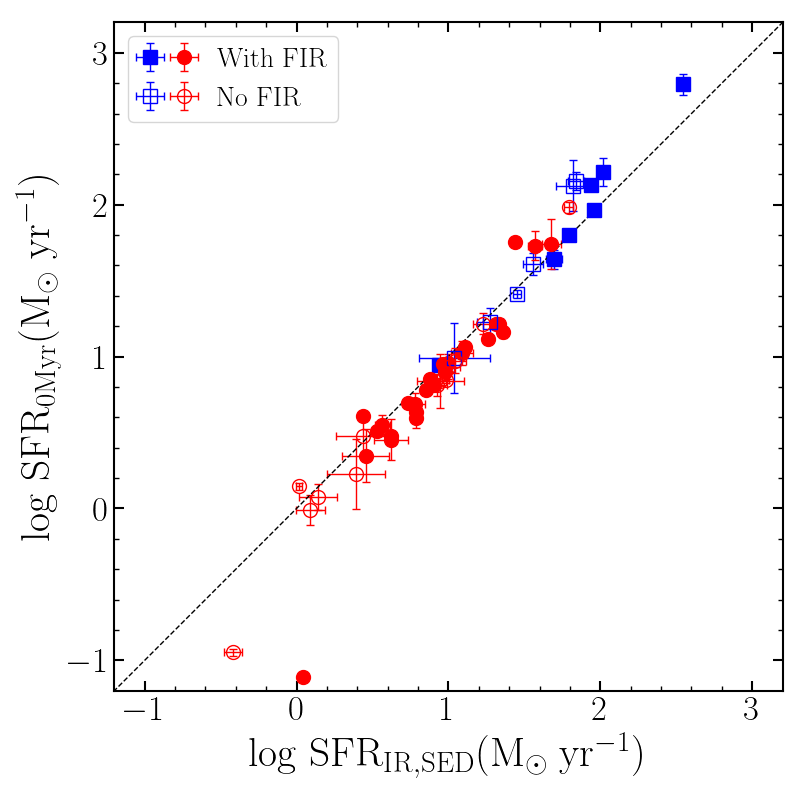}
\includegraphics[width=0.5\textwidth]{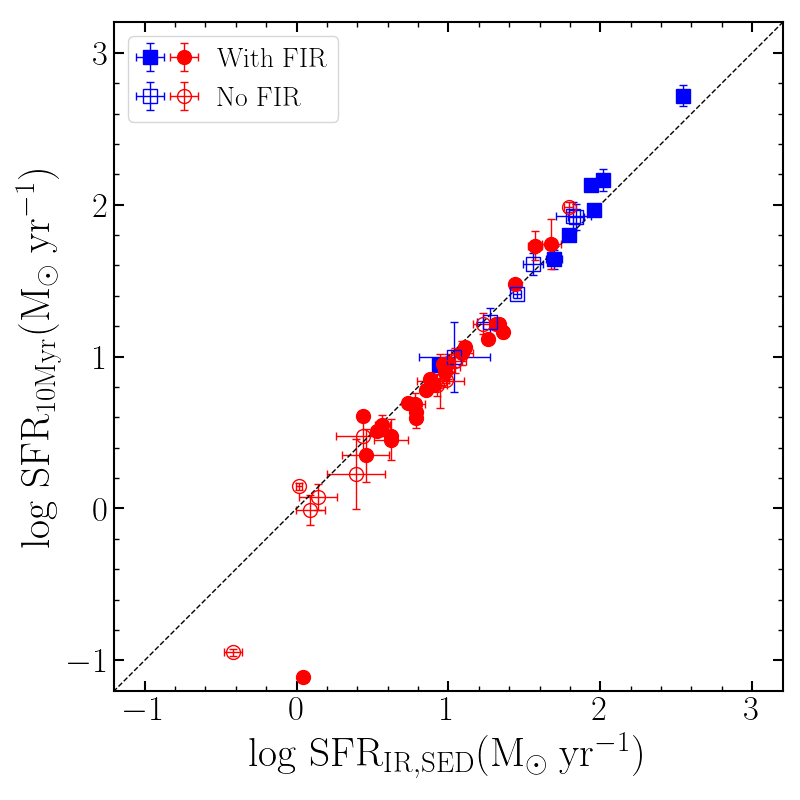}
\caption{Comparison of the SFR based on dust luminosity with the instantaneous SFR (left) or average SFR over 10 Myr (right) estimated from stellar population models. Type 1 and type 2 targets are represented with blue squares and red circles, respectively. Open symbols denote targets with no FIR or sub-mm detection at $\lambda >30\;\mu$m.}
\label{SFH_SFR_comparison}
\end{figure*}

Although we adopt dust luminosity based SFR in this study to minimize the uncertainty induced by AGN contamination, CIGALE also provides several SFRs calculated from star formation history models. Here we compare the dust luminosity based SFRs with SFR estimates from star formation history models, to check their consistency. Three SFRs are provided by CIGALE, namely, instantaneous SFR (sfh.sfr) and average SFRs over 10 Myr or 100 Myr, respectively.
Since dust luminosity based SFR is thought to trace mainly $\sim$5 Myr age of stellar population over up to a 10$^{8}$ yrs time scale \citep{Kennicutt+12}, we compare dust luminosity SFR with instantaneous SFR and average SFR over 10 Myrs, respectively in Figure \ref{SFH_SFR_comparison}.

We find that dust luminosity based SFR shows good consistency with stellar population model based SFRs in general. Instantaneous SFR (i.e., SFR$_{\mathrm{0Myr}}$) is slightly lower than dust luminosity based SFR by a 0.04 dex offset with a 0.22 dex scatter. Average SFR over 10 Myr has a similar trend, as it is lower by a 0.06 dex offset with a 0.20 dex scatter compared to dust luminosity based SFR. Note that there are two outliers at the low end of SFR range, of which star formation is mostly quenched without any young stellar population (i.e., lack of recent star formation episode; see Table \ref{table_CIGALEpar}).
Therefore, the relatively large contribution of old stars to dust luminosity is presumably responsible for the overestimation of SFR based on dust luminosity in these targets. Albeit with the two outliers, the overall trends are consistent with a one-to-one relation within 0.20 dex, indicating dust luminosity based SFR is consistent with stellar population model based SFR.

\subsection{Effect of AGN dust heating}

\begin{figure}[h]
	\includegraphics[width=\columnwidth]{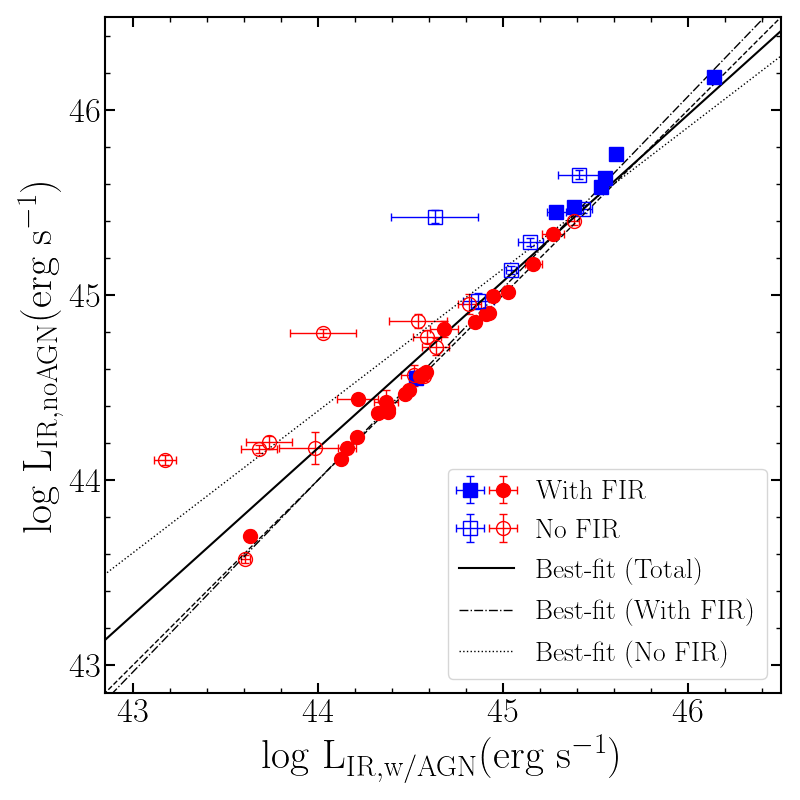}
	\caption{Comparison of dust IR luminosities obtained with and without including an AGN dust component in the SED fitting. The best-fit is represented by a solid line for the total sample, while the best-fits for targets with and without FIR detection are denoted with a dash dotted and dotted line, respectively. A one-to-one relationship is denoted by a dashed line. Open symbols indicate targets with only one detection in MIR at $\lambda <30\;\mu$m.}
	\label{agnnoagn}
\end{figure}

We test the effect of AGN dust heating on the SED analysis and dust luminosity determination. Since CIGALE allows us to determine the best-fit SED model with and without including the AGN component, which is dominant in the MIR range, we can constrain the AGN contribution in the total IR luminosity by comparing the SED fitting results. To find the best-fit SED without including the AGN hot dusty torus component, we used the same method as described in Section \ref{subsec:SED}, while we fixed `fracAGN' parameter as 0.

In Figure 8, we compare the dust luminosity measured from the SED fitting with and without including the AGN component. Overall, the dust luminosity is slightly overestimated by 0.13 dex and there is a 0.25 dex scatter. We find a best-fit slope of 0.90$\pm$0.04 for the sample. 
We also divide the sample into two subsamples based on the availability of FIR or sub-mm data. In general we find a larger discrepancy if FIR or sub-mm data are unavailable. For example, the subsample of targets with only MIR points (i.e., $\lambda < 30\;\mu$m) in the IR range shows a 0.26 dex offset and a 0.39 dex scatter, while the subsample of targets with FIR or sub-mm data have a much smaller offset and a scatter, i.e., 0.04 and 0.07 dex, respectively. Also, the best-fit slope of the targets with FIR or sub-mm data (dash dotted line) is almost a one-to-one relation ($1.04\pm0.01$), while the subsample of targets without FIR nor sub-mm data shows a systematic trend with a slope of $0.77\pm0.09$ (dotted line).

We find that FIR ($\lambda > 30\;\mu$m) data, as well as sub-mm data in the absence of FIR, play an important role in obtaining a good SED model of the dust component as similarly shown in Section \ref{subsec:CE01comparison}.
It was previously reported that the availability of FIR and sub-mm data improves SFR measurements, particularly for AGN host galaxies, and that SFR can be significantly overestimated if AGN hot dusty torus component is not included in the SED analysis \citep{Ciesla:2015A&A, Florez+20}. Our results clearly show the same case.

We also investigate the dependency of the AGN type by separating type 1 and type 2 AGNs. We find 0.15 dex and 0.12 dex offsets, respectively for type 1 and type 2 AGNs, while the scatter is comparable (0.25 dex) to each other. Thus, we find no clear difference between type 1 and type 2 AGN in determining the dust luminosity.

In summary, the effect of AGN dust heating on derived IR luminosity is relatively small, but not negligible for our sample of low-z AGN host galaxies. Especially, when FIR data are unavailable, the derived IR luminosity can be overestimated by a factor of two.

\subsection{Comparison of SFRs}\label{comp_SFRs}

\begin{figure}[h]
	\includegraphics[width=\columnwidth]{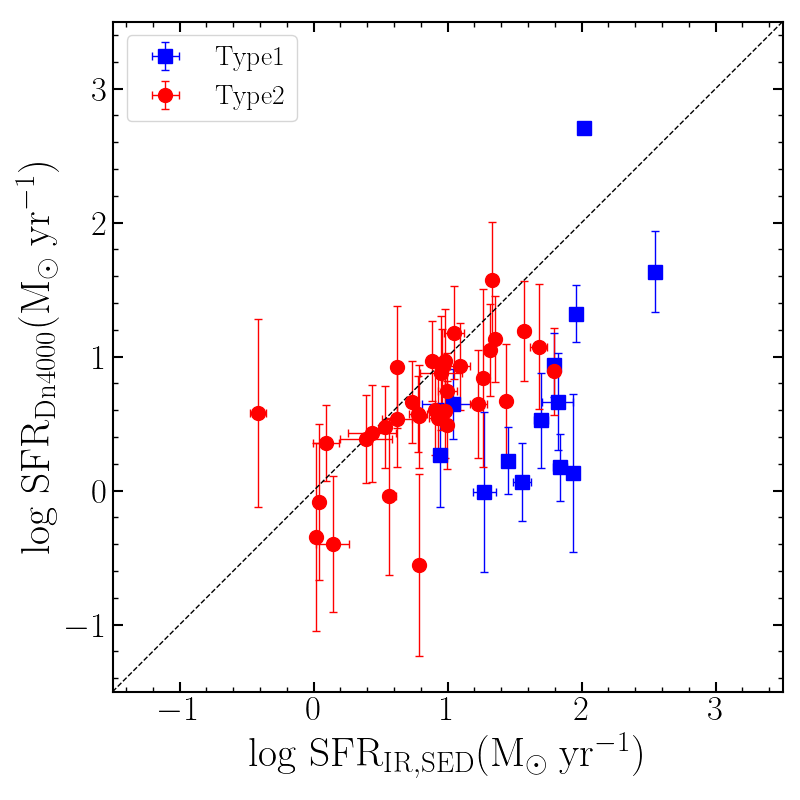}

	\caption{Comparison of the SFR based on dust luminosity obtained from CIGALE with the SFR obtained from D$_{n}$4000 break for type 1 (blue squares) and type 2 AGNs (red circles).
	}
	\label{4000}
\end{figure}

\begin{figure}[h]
\includegraphics[width=\columnwidth]{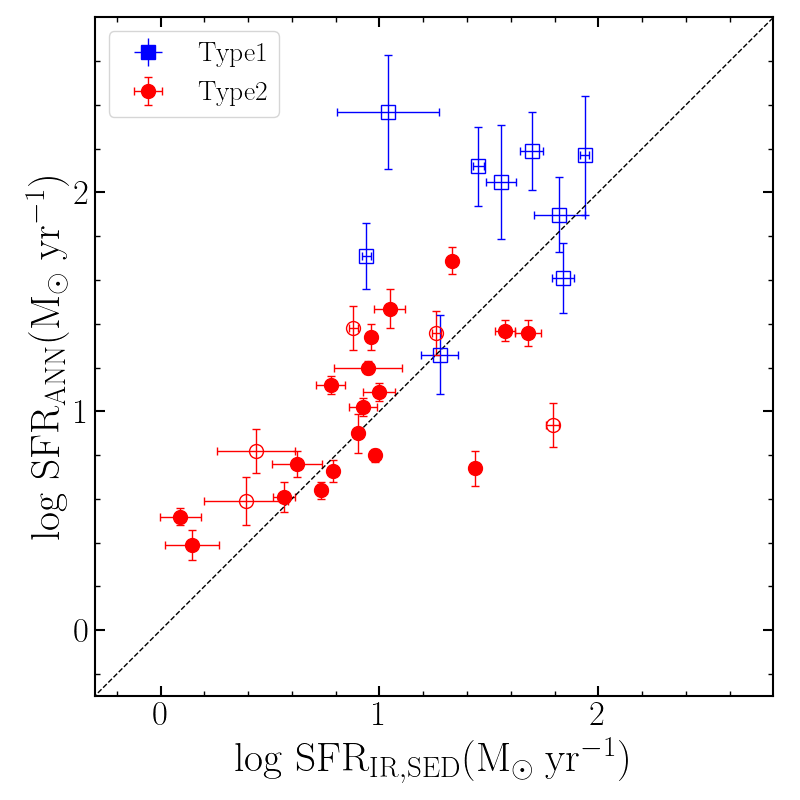}

\caption{Comparison of the SFR based on dust luminosity obtained from CIGALE with the ANN-based SFR, for type 1 (blue squares) and type 2 AGNs (red circles). Open symbols indicate targets with $\sigma_{ANN}>0.1$, while filled symbols indicate $\sigma_{ANN}<0.1$.
}
\label{ann}
\end{figure}

While we determine the SFR based on the dust luminosity measured from SED fitting, other SFR estimates are available for a subsample of our targets. Although it is difficult to measure SFR of AGN host galaxies, two estimates are available and we present the comparison results in this section. 

First, we utilize the SFR calculated from the 4000\AA\ break provided by the MPA-JHU value-added catalog\footnote{http://www.mpa-garching.mpg.de/SDSS/}.  The flux difference around 4000\AA\ (D$_{n}$4000) traces the age of stellar populations, hence, SFR is derived from the measured D$_{n}$4000 as outlined by \citet{Brinchmann:2004MNRAS}. 
We compare the SFR determined from the dust luminosity based on our SED fitting analysis (SFR$_{\rm{IR, SED}}$) with the SFR based on 4000\AA$\;$break (SFR$_{\rm{Dn4000}}$) in Figure \ref{4000}.

On average, the SFR$_{\rm{IR, SED}}$ is higher than the SFR$_{\rm{Dn4000}}$ by 0.43 dex with a scatter of 0.71 dex, indicating that the D$_{n}$4000 method underestimates SFR compared to the dust luminosity method. In particular, type 1 AGNs (a 0.97 dex offset and a 1.15 dex scatter) clearly show a much larger offset and a scatter than type 2 (a 0.23 dex offset and a 0.46 dex scatter). 

For AGN host galaxies, D$_{n}$4000-based SFRs may not be reliable due to the contamination of AGN continuum and line emission and this systematic effect may cause the significant offsets compared to the dust luminosity based SFR \citep[e.g.,][]{Rosario16}. Note that we corrected $\sim 6\%$ for SFR$_{\rm{Dn4000}}$ values to compare the two SFRs in a consistent manner with the same Chabrier IMF \citep{Madau14}.

Second, we compare the dust luminosity based SFR with the SFR estimate based on Artificial Neural Network (ANN) analysis by \citet{Ellison16a}. Using a training sample of galaxies with well-defined SFRs, optical properties including emission line strength have been mainly used to constrain the total IR luminosity. While the uncertainty of SFR for individual target can be significant, these SFR estimates are available for a large sample of the SDSS AGN host galaxies and useful for statistical stidues \citep[e.g.,][]{Ellison16a, Woo:2017ApJ, Woo:2020}. Among 52 AGNs in our sample, the ANN-based SFR (SFR$_{\rm{ANN}}$) is available for 23 type 2 AGNs and 9 type 1 AGNs. Thus, we present the comparison of the two SFR for 32 AGNs in Figure \ref{ann}.

We find an average offset of 0.17 dex, indicating that SFR$_{\rm{ANN}}$ is slightly larger than but comparable to SFR$_{\rm{IR, SED}}$, while there is a substantial scatter of 0.44 dex. However, if we focus on the targets with a good ANN-based SFR (i.e., small ANN uncertainty $\sigma_{ANN}<0.1$), an offset and scatter significantly decrease to 0.07 dex and 0.30 dex, respectively, indicating a consistency between the two SFRs.
There is a clear distinction of the systematic offset between type 1 and type 2 AGNs. While SFR$_{\rm{ANN}}$ is larger than SFR$_{\rm{IR, SED}}$  by 0.42 dex with a 0.62 dex scatter for type 1 AGNs, SFR$_{\rm{ANN}}$ of type 2 AGNs is comparable to SFR$_{\rm{IR, SED}}$ with a 0.07 dex offset and a 0.35 dex scatter. It is not clear why SFR$_{\rm{ANN}}$ of type 1 AGNs is overestimated compared to the dust luminosity based SFR. Presumably, the contribution of the broad emission lines and continuum may affect the line flux measurements which were used in the ANN analysis.
Our results suggest that while ANN-based SFR is not reliable for individual objects as expected, SFR$_{\rm{ANN}}$ provides valuable constraints for statistical studies if we limit the sample with high quality ANN estimates.

\subsection{Correlation of SFR with AGN outflows}\label{corr_SFR_outflow}

\begin{figure}[h]
\includegraphics[width=\columnwidth]{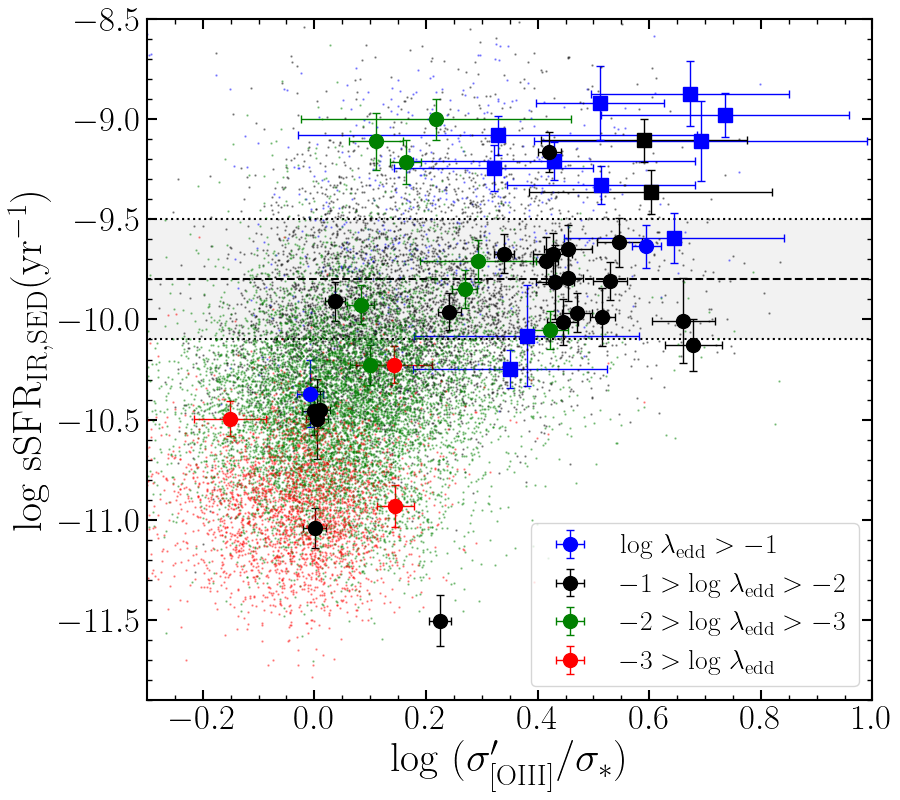}

\caption{Comparison of sSFR with the normalized [\OIII]\ velocity dispersion for type 2 (circles) and type 1 AGNs (squares) as well as SDSS type 2 AGNs (dots) adopted from \citet{Woo:2017ApJ}. AGN are divided into four bins according to Eddington ratios. Note that all type 1 AGNs have an Eddington ratio larger than 1\%.
The main sequence of star-forming galaxies is represented by a dashed line with a 1$\sigma$ dispersion as measured for non-AGN galaxies by \cite{Woo:2017ApJ}.
}
\label{sSFRoutflow}
\end{figure}

\begin{figure}[h]
\includegraphics[width=\columnwidth]{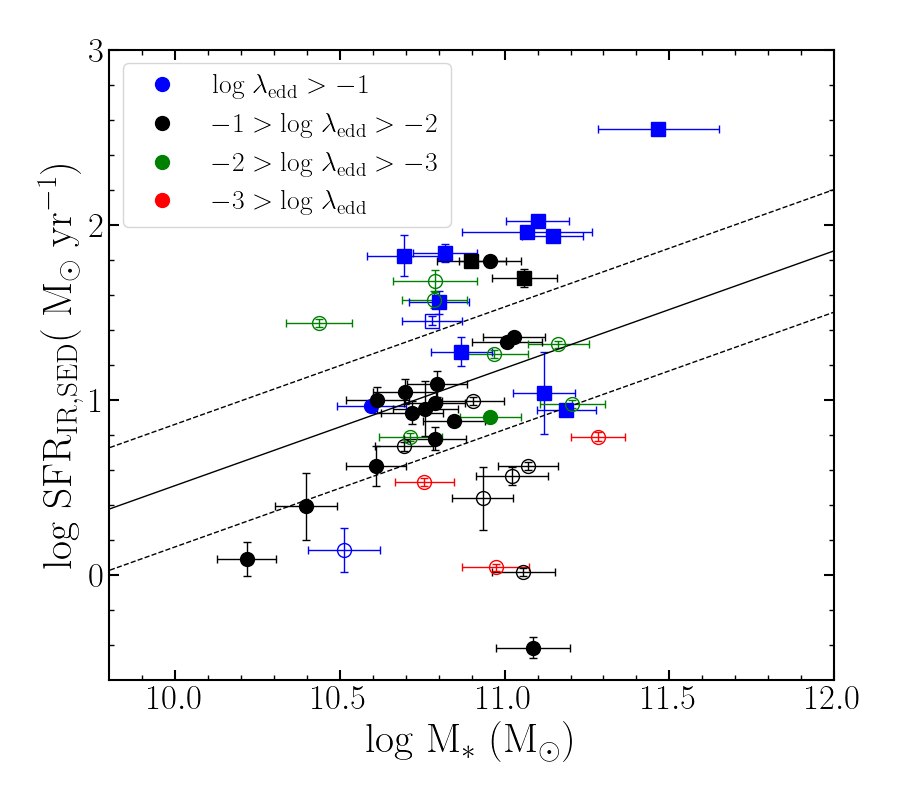}
\caption{SFR vs. stellar mass for type 2 (circles) and type 1 AGNs (squares).  The solid line denotes the main sequence 
along with the $1\sigma$ dispersion adopted from \citet{Noeske:2007ApJ}. AGN are divided into four bins according to Eddington ratios. Note that AGN with strong outflows are represented by filled symbols while AGNs with low outflows are denoted with open symbols.
}
\label{sfrmstar}
\end{figure}

Using the best SFR obtained based on dust luminosity, we investigate the correlation between SFR and the strength of ionized gas outflows to test whether our sample also follows the same correlation reported by \citet{Woo:2020}, who demonstrated that AGNs with stronger outflows tend to be hosted by galaxies with higher SFRs using a large sample of SDSS AGNs \citep[see also][]{Woo:2017ApJ, Luo+21}. While the previous studies were performed based on SFR estimates from the monochromatic luminosity measured at FIR or the ANN analysis, we can directly compare the dust luminosity based SFR (SFR$_{\rm{IR, SED}}$) with outflow properties, using our SED sample. 

As a tracer of outflow kinematics, we use [\OIII] velocity dispersion from our previous studies \citep{Woo+16,Rakshit18}, which presented a demography of ionized gas
outflows using a large sample of 39,000 type 2 AGNs and 5,000 type 1 AGNs at z$<$0.3. In Figure \ref{sSFRoutflow}, we compare outflow kinematics with specific star formation rate (sSFR), which is determined by normalizing SFR$_{\rm{IR, SED}}$ by stellar mass. We also plotted SDSS type 2 AGNs from \citet{Woo:2017ApJ}, for which SFR is available based on FIR flux measurements or ANN method.
Note that [\OIII]\ velocity dispersion is also normalized by stellar velocity dispersion to account for the effect of host galaxy gravitational potential as similarly performed in the previous studies \citep{Karouzos+16a, Woo+16, Woo:2017ApJ, Luo+19, Woo:2020}. Thus, if ionized gas velocity dispersion is much larger than stellar velocity dispersion, we interpret that gas outflows are very strong. 

We find a broad trend that stronger outflow AGNs tend to be hosted by higher SFR galaxies. Note that since our SED sample is composed of luminous AGNs with and without strong outflows (see Figure \ref{o3dispo3vel}), the range of the normalized [\OIII]\ velocity dispersion of the SED sample is broad, covering that of the SDSS sample, while a large fraction of the SED sample shows strong outflow kinematics, i.e., [\OIII]\ velocity dispersion is larger than stellar velocity dispersion by a factor of two.

In order to investigate whether SFR is related with AGN energetics, we divide our SED sample into four bins according to Eddington ratios, as similarly analyzed by \cite{Woo:2020}.
There is a general trend that most of the AGNs with relatively high Eddington ratios (i.e., $>$1\%) have high outflow velocities, whereas most of low Eddington ratio AGNs ($<$ 1\%) have lower outflow velocities  \citep[for a quantitative result, see Table 2 and Figure 5 of][]{Woo+16}. At the same time, SFR broadly correlates with Eddington ratio, following the general trends shown by the SDSS sample \citep{Woo:2017ApJ, Woo:2020}.

These results indicate that when AGN activity is strong (i.e., high Eddington ratio and strong outflows), star formation activity is also strong. As reported by \citet{Woo:2020}, the positive correlation of outflow strength with SFR suggests lack of instantaneous quenching of star formation. Our SED sample with well measured SFR based on dust luminosity supports this scenario.  
We point out that there could be a systematic difference of Eddington ratio determination between type 1 and type 2 AGNs since the black hole mass of type 2 AGNs suffer from large systematic uncertainties as stellar mass along with the black hole mass - stellar mass relation were used to derive black hole mass \citep[for details, see][]{Woo:2017ApJ}. Also, AGN bolometric luminosity has a larger uncertainty for type 2 AGNs as it was estimated from the [\OIII]\ emission line luminosity. Thus, the Eddington ratio of type 2 AGNs is much less reliable compared to type 1 AGNs, for which H$\beta$ emission line velocity dispersion and the continuum luminosity at 5100\AA\ were used to calculate black hole mass, and the 5100\AA\ luminosity was used to determine AGN bolometric luminosity \citep[for details, see][]{Rakshit18}. The reason why type 2 AGNs show a less clear trend in Figure \ref{sSFRoutflow} is presumably due to these uncertainties. 

In Figure \ref{sfrmstar} we compare the SFR based on dust luminosity with stellar mass. The star-forming main sequence is represented by a gray line with a 1 $\sigma$ dispersion, which is adopted from \citet{Noeske:2007ApJ} as log$(SFR) = (0.67 \pm 0.08)$ log$(M_\star) - (6.19 \pm 0.78)$. A larger fraction of our SED sample lies within the 1$\sigma$ distribution of the main-sequence galaxies, albeit with a relatively broad distribution of SFR for given stellar mass. 
As shown in Figure \ref{sSFRoutflow}, we separate our SED sample into 4 different Eddington ratio bins, finding a broad trend that higher Eddington ratio AGNs tend to have higher SFR at given stellar mass (for the correlation between SFR and Eddington ratio, see Section \ref{corr_SFR_ER}).
While SFRs of the targets with moderate Eddington ratios (black and green points) are distributed too broad to see a clear difference between them, there is a general trend that SFRs of the highest Eddington ratio AGNs (blue points) are higher or similar to main-sequence SFR level except only one. On the contrary, AGNs with the lowest Eddington ratio (red points) are all located below the main-sequence. Thus, the broad trend is consistent with the findings reported by \citet{Woo:2020}. These results suggest that high Eddington ratio AGNs generally show strong outflows as well as high star formation rate, indicating no instantaneous quenching of star formation. 

\subsection{Correlation of SFR with Eddington ratio}\label{corr_SFR_ER}

\begin{figure}[h]
\centering
\includegraphics[width=\columnwidth]{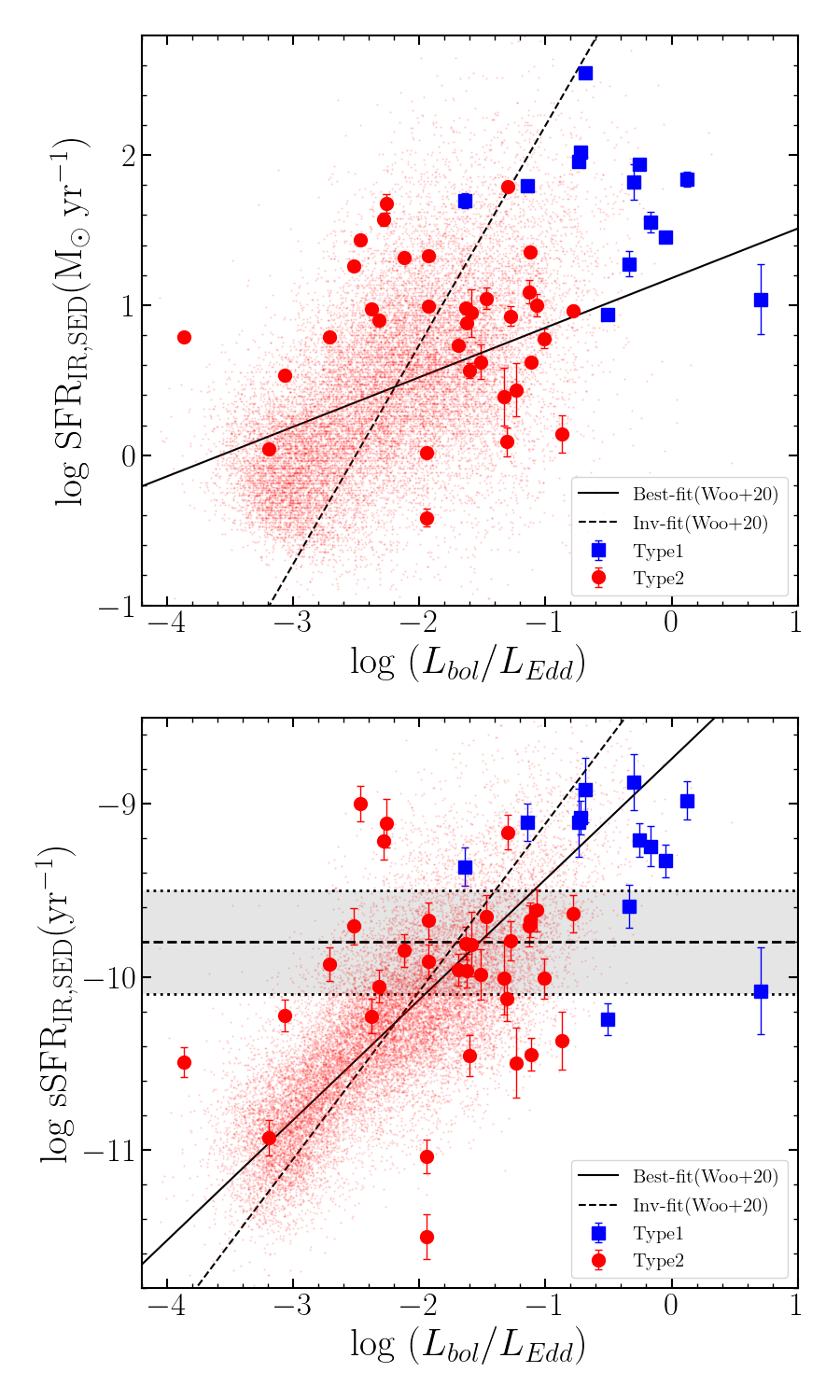}
\caption{SFR vs. Eddington ratio (top) and sSFR vs. Eddington ratio (bottom). Grey area in the below figure represents 1$\sigma$ distribution of sSFR of main sequence galaxies \citep{Woo:2017ApJ}. SDSS type 2 AGNs from \citet{Woo:2017ApJ} are shown as red dots. Solid lines and dashed lines represent best and inverse fits from \citet{Woo:2020}.
}
\label{ER_SFR}
\end{figure}

We compare SFR or sSFR with Eddington ratio of the SED sample in Figure \ref{ER_SFR}. As \citet{Woo:2020} reported a positive correlation of SFR (sSFR) with Eddington ratio by a slope of $0.33\pm0.02$ ($0.69\pm0.02$) based on a large sample of type 1 and type 2 AGNs, we find a general positive trend between SFR or sSFR with Eddington ratio. We perform correlation analysis by significance tests and find that both SFR and sSFR correlate with Eddington ratio by p-value $\lesssim 0.01$. The tighter correlation of Eddington ratio with sSFR than SFR is also consistent with \citet{Woo:2020}.
On the other hand, scatters are relatively large ($\gtrsim$ 0.55 dex) and the best-fit slopes of the SED sample are much shallower than those of \citet{Woo:2020}. These may originate from the sample selection, which was focused on strong outflow signatures (i.e., high Eddington ratio) and/or the small sample size. Nevertheless, our results show no significant disagreements with previous trends. Further studies with reliable SFR for a larger sample may better constrain the nature of the correlation.

\section{Summary \& Conclusions}\label{sec:con}

We presented sub-mm observations and measurements with the JCMT SCUBA-2 for a sample of 52 AGN host galaxies at z $<0.2$. Combining with archival multi-wavelength data, we fit the SED of individual targets to determine dust luminosity and SFR. We investigated the effect of AGN dust heating, reliability of other SFR estimates, and the relation with AGN activity.
Our main results are summarized as follows. 

\begin{itemize}
\item While dust IR luminosity can be contaminated by AGN hot dusty torus emission, the effect is relatively small for low-z AGN host galaxies ($\sim$30 - 40 \%). However, it is substantially larger by a factor of $\sim$two when there is no FIR or sub-mm data ($\lambda > 30\;\mu$m) in the SED modeling.

\item 
Comparing with the dust luminosity based SFR, we find that SFR based on D$_n$4000 shows a significant offset, while the ANN-based SFR is comparable, particularly when the sample is limited with high quality ANN estimates ($\sigma_{ANN}<0.1$). In general type 1 AGNs show a more significant offset than type 2 AGNs, presumably due to the contamination of additional AGN emission from an accretion disk and the broad line region in D$_n$4000 measurements and ANN analysis. 

\item We find a positive trend of SFR with AGN outflow strength represented by [\OIII]\ velocity dispersion and Eddington ratio, confirming the correlation reported by \citet{Woo:2017ApJ} and \citet{Woo:2020}. The general trend that high Eddington ratio AGNs with strong outflows are hosted by high-SFR galaxies suggest that there is lack of instantaneous quenching of star formation due to AGN feedback. 

\end{itemize}

\acknowledgements{We thank the anonymous referee for valuable comments, which improved the clarity of the manuscript.
This work has been supported by the Basic Science Research Program through the National Research Foundation of Korean Government (NRF-2021R1A2C3008486). We would like to thank Marios Karouzos for his valuable input for the JCMT observation project. HSH was supported by the New Faculty Startup Fund from Seoul National University.}

\software{Astropy \citep{Astropy1, Astropy2}, CIGALE \citep{Noll:2009A&A, Boquien19}, CUPID \citep{Berry+07, Berry+15}, matplotlib \citep{Hunter07_matplotlib}, numpy \citep{Harris20_Numpy}, ORAC-DR \citep{Jenness+15}, pandas \citep{McKinney10_pandas}, scipy \citep{Virtanen20_Scipy}, SMURF \citep{Chapin+13_Smurf}, Starlink \citep{Currie+14_Starlink}}

\appendix
\section{Mock Catalog Analysis}
\begin{figure*}[]

\includegraphics[width=0.5\textwidth]{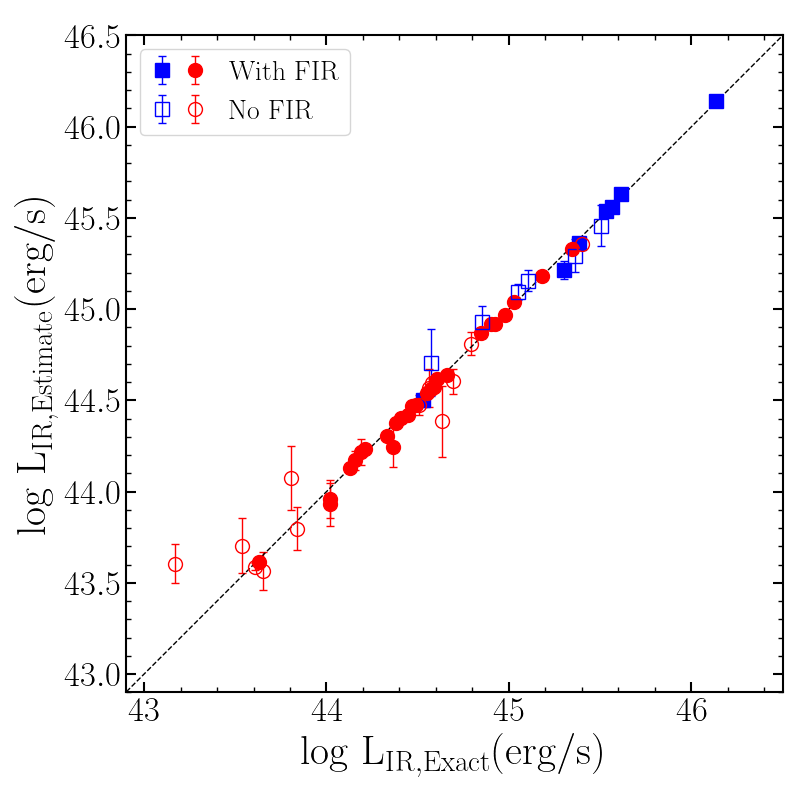}
\includegraphics[width=0.5\textwidth]{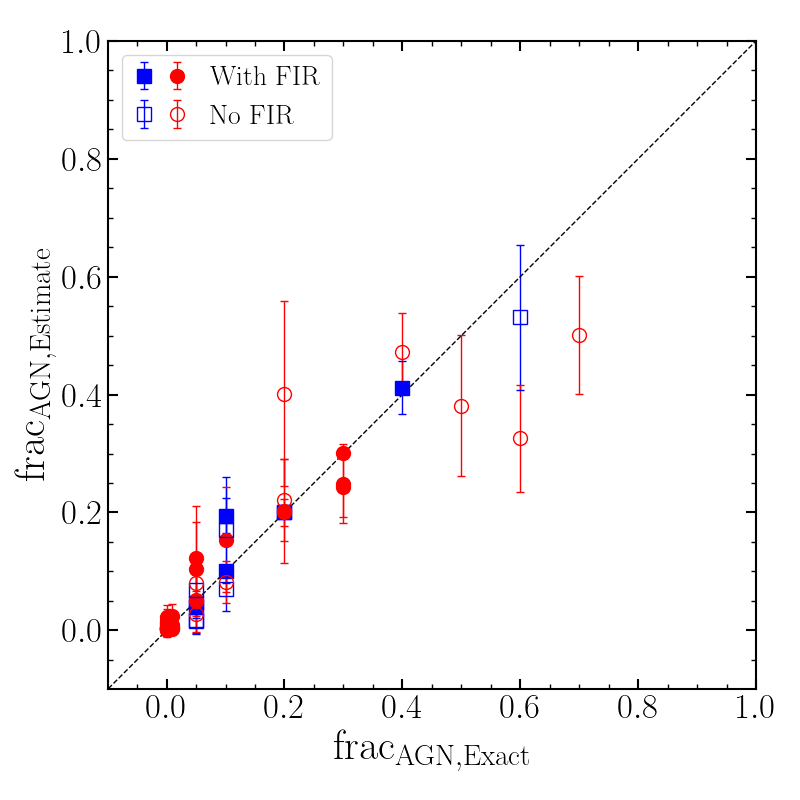}
\caption{Comparison between exact parameter values of the best-fit mock SEDs and the estimated values by CIGALE SED fitting. The left and right panel show comparison plots of dust IR luminosity and AGN fraction parameters, respectively. A dashed line indicates a one-to-one relation. Open symbols denote targets with no FIR or sub-mm detection at $\lambda > 30\;\mu$m.}
\label{mock_analysis}
\end{figure*}

We investigated the reliability of the dust luminosity and AGN fraction estimates from the CIGALE SED fitting, using the mock catalog analysis provided by
CIGALE. This process builds a Gaussian distribution for each filter using an extracted flux from the best-fit SED and an observational error. Then it randomly generates mock observations according to the Gaussian distributions and re-performs Bayesian SED fitting analysis. Thus, we can compare the known parameters with the estimated parameters from the re-performed Bayesian analysis \citep[for details, see][and CIGALE manual]{Boquien19}.

As shown in Figure \ref{mock_analysis} (left panel), we find that the Bayesian estimates of IR luminosity (L$_{\rm IR, Estimate}$) are  consistent with the known IR luminosity (L$_{\rm IR, Exact}$) with no clear offset and a 0.09 dex scatter. If we limit the targets with available FIR or sub-mm data, the scatter becomes negligible (0.03 dex), indicating that CIGALE reproduces the intrinsic IR luminosity without significant systematic uncertainties.
In the case of AGN fraction (right panel), we also find that the Bayesian estimates are consistent with the input values. After removing targets with zero AGN fraction, we calculated the scatter in the logarithmic scale, which is 0.14 dex. For the targets with more reliable SED fitting results with FIR or sub-mm data, the scatter is 0.13 dex, indicating no strong systematic uncertainties.

\bibliography{ref}

\end{document}